\documentclass[12pt]{article}
\usepackage{graphics,epsfig,cite,psfrag,amsmath,longtable}

\def\citere#1{\mbox{Ref.~\cite{#1}}}

\def\Mfunction#1{\displaystyle #1}

\def\Mvariable#1{\text{#1}}

\def\i{\mathrm{i}}
\def\tiu{\tilde u}
\def\tid{\tilde d}
\def\tiq{\tilde q}
\def\De{\Delta}
\def\order#1{${\cal O}(#1)$}
\def\al{\alpha}
\def\be{\beta}
\newcommand{\cp}{{\cal CP}}

\newcommand{\oaas}{{\cal O}(\alpha\alpha_s)}

\newcommand{\twol}{two-loop}
\newcommand{\onel}{one-loop}

\newcommand{\fa}{{\em FeynArts}}

\newcommand{\fhto}{{\em FeynHiggs2.1}}

\newcommand{\MW}{M_W}
\newcommand{\MZ}{M_Z}

\newcommand{\MA}{M_A}
\newcommand{\mh}{m_h}

\newcommand{\mhtree}{m_{h,{\rm tree}}}
\newcommand{\mHtree}{m_{H,{\rm tree}}}
\newcommand{\msusy}{M_{\SU}}
\newcommand{\msq}{m_{\tilde{q}}}
\newcommand{\mhmax}{\mh^{\rm max}}

\newcommand{\sw}{s_W}
\newcommand{\cw}{c_W}
\newcommand{\sweff}{\sin^2\theta_{\mathrm{eff}}}
\newcommand{\tev}{\,\, \mathrm{TeV}}
\newcommand{\gev}{\,\, \mathrm{GeV}}
\newcommand{\mev}{\,\, \mathrm{MeV}}
\def\de{\delta}
\def\Si{\Sigma}
\def\hSi{\hat{\Sigma}}
\def\aeff{\al_{\rm eff}}
\newcommand{\SU}{\mathrm {SUSY}}

\newcommand{\Stop}{\tilde{t}}
\newcommand{\Scha}{\tilde{c}}
\newcommand{\Sbot}{\tilde{b}}
\newcommand{\Sstr}{\tilde{s}}
\newcommand{\StopL}{\tilde{t}_L}
\newcommand{\StopR}{\tilde{t}_R}
\newcommand{\SchaL}{\tilde{c}_L}
\newcommand{\SchaR}{\tilde{c}_R}
\newcommand{\SstrL}{\tilde{s}_L}
\newcommand{\SstrR}{\tilde{s}_R}
\newcommand{\SbotL}{\tilde{b}_L}
\newcommand{\SbotR}{\tilde{b}_R}
\newcommand{\mt}{m_{t}}

\newcommand{\mq}{m_{q}}
\newcommand{\mb}{m_{b}}
\newcommand{\mc}{m_c}
\newcommand{\ms}{m_s}
\newcommand{\At}{A_t}

\newcommand{\Aq}{A_q}

\newcommand{\Xt}{X_t}
\newcommand{\Xb}{X_b}

\newcommand{\Xc}{X_c}
\newcommand{\Xs}{X_s}

\newcommand{\sfl}{\tilde{f}_L}
\newcommand{\sfr}{\tilde{f}_R}
\def\refse#1{\mbox{Sect.~\ref{#1}}}

\def\reffi#1{\mbox{Fig.~\ref{#1}}}

\def\la{\lambda}

\def\ga{\gamma}
\newcommand{\lsim}
{\;\raisebox{-.3em}{$\stackrel{\displaystyle <}{\sim}$}\;}
\newcommand{\gsim}
{\;\raisebox{-.3em}{$\stackrel{\displaystyle >}{\sim}$}\;}
\newcommand{\BE}{\begin{equation}}
\newcommand{\BEA}{\begin{eqnarray}}
\newcommand{\BEAnn}{\begin{eqnarray*}}
\newcommand{\EEA}{\end{eqnarray}}
\newcommand{\EEAnn}{\end{eqnarray*}}
\newcommand{\VL}{\left( \begin{array}{c}}
\newcommand{\VR}{\end{array} \right)}
\newcommand{\KL}{\left(}
\newcommand{\KKL}{\left[}
\newcommand{\KR}{\right)}
\newcommand{\KKR}{\right]}
\newcommand{\non}{\nonumber}
\newcommand{\MLv}{\left( \begin{array}{cccc}}
\newcommand{\MR}{\end{array} \right)}
\newcommand{\ML}{\left( \begin{array}{cc}}
\newcommand{\MLd}{\left( \begin{array}{ccc}}
\newcommand{\KKKL}{\left\{}
\newcommand{\KKKR}{\right\}}
\newcommand{\tb}{\tan \beta}
\newcommand{\sa}{s_\al}
\newcommand{\ca}{c_\al}
\newcommand{\Sa}{\sin \alpha\hspace{1mm}}
\newcommand{\Ca}{\cos \alpha\hspace{1mm}}
\newcommand{\cbe}{c_\be}
\newcommand{\sbe}{s_\be}

\newcommand{\CZb}{\cos 2\beta\hspace{1mm}}
\def\_{\rule{.3em}{.15ex}} 
\newcommand{\feynmancaption}[1]{\addtocounter{subsection}{1}%
\\ {\bf\Large \thesubsection{} \ #1}\\*[3ex]
\addcontentsline{toc}{subsection}{\thesubsection{} \ #1}%
}
\newcommand{\nn}{\nonumber}

\def\slash#1{\setbox0=\hbox{$#1$}#1\hskip-\wd0\dimen0=5pt\advance
       \dimen0 by-\ht0\advance\dimen0 by\dp0\lower0.5\dimen0\hbox
         to\wd0{\hss\sl/\/\hss}}

\def\swq{s_W^2}
\def\cwq{c_W^2}

\def\eps{$\epsilon$ }
\def\lann{$\lambda \neq 0$ }

\allowdisplaybreaks

\begin{document}
\thispagestyle{empty}

\def\thefootnote{\fnsymbol{footnote}}

\begin{flushright}
CERN--PH--TH/2004--53\\
MPP--2004--31\\
hep-ph/0403228 \\
\end{flushright}

\vspace{1cm}

\begin{center}
{\large\sc {\bf Electroweak Precision Observables in the MSSM}}
\vspace{0.4cm}
{\large\sc {\bf with Non-Minimal Flavor Violation}}
\vspace{1cm}

{\sc 
S.~Heinemeyer$^{1}$%
\footnote{email: Sven.Heinemeyer@cern.ch}%
, W. Hollik$^{2}$%
\footnote{email: hollik@mppmu.mpg.de}%
, F. Merz$^{2}$%
\footnote{email: merz@mppmu.mpg.de}%
~and~S.~Pe\~naranda$^{2}$%
\footnote{email: siannah@mppmu.mpg.de}
}

\vspace*{1cm}

{\sl
$^1$CERN TH Division, Department of Physics,\\ 
CH-1211 Geneva 23, Switzerland

\vspace*{0.4cm}

$^2$Max-Planck-Institut f\"ur Physik (Werner-Heisenberg-Institut),\\
F\"ohringer Ring 6, D--80805 Munich, Germany
}

\end{center}

\vspace*{0.2cm}

\begin{abstract}
The leading corrections to electroweak precision observables in the
MSSM with non-minimal flavor violation (NMFV) are calculated and the
effects on $\MW$ and $\sweff$ are analyzed.
The corrections are obtained by evaluating the full \onel\ contributions 
from the third and second generation scalar quarks, including the
mixing in the scalar 
top and charm, as well as in the scalar bottom and strange sector.
Furthermore the leading corrections to the mass of the lightest MSSM
Higgs boson, $\mh$, is obtained. The electroweak one-loop contribution 
to $\MW$ can amount up to $140 \mev$ and up
to $70 \times 10^{-5}$ for $\sweff$, allowing to set limits on the
NMFV parameters. The corrections for $\mh$ are not significant for
moderate generation mixing. 
\end{abstract}

\def\thefootnote{\arabic{footnote}}
\setcounter{page}{0}
\setcounter{footnote}{0}

\newpage

\section{Introduction}

Supersymmetric theories of the strong and electroweak interactions, like
the Minimal Supersymmetric Standard Model (MSSM)~\cite{susy}
as the theoretically favored extension of the Standard Model (SM),
predict the existence of scalar partners $\tilde{f}_L, 
\tilde{f}_R$ to each SM chiral fermion, and of spin-1/2 partners to the 
gauge and Higgs bosons. So far, the direct search for
SUSY particles could only set lower bounds of ${\cal O}(100)$~GeV on 
their masses~\cite{pdg}. In a similar way, the search for MSSM Higgs bosons  
resulted in lower limits of about $90 \gev$ for the
neutral and  $80 \gev$ for the charged Higgs particles~\cite{LEPHiggs}. 

An alternative way, as compared to the direct search for SUSY or Higgs
particles, is to probe SUSY via virtual effects of the 
additional non-standard particles to precision observables. 
This requires very high precision 
of the experimental results as well as of the theoretical predictions.
A predominant role in this respect has to be assigned to the
$\rho$-parameter~\cite{rho}, with loop contributions $\De\rho$
through vector-boson self-energies constituting the leading 
process-independent quantum corrections 
to electroweak precision observables, such as the prediction for
$\De r$ in the $\MW$--$\MZ$~interdependence 
and the effective leptonic weak mixing angle, $\sweff$. 

Radiative corrections to the electroweak precision observables
within the MSSM, originating from the virtual presence of
scalar fermions, charginos, neutralinos, and Higgs bosons, 
have been discussed at the \onel\ level in~\cite{dr1lA,dr1lB},
providing the full \onel\ corrections.
More recently, also the leading \twol\ contributions in $\oaas$
to $\De\rho$ from quarks, squarks, gluons, and gluinos  
have been obtained~\cite{dr2lA} as well as the gluonic \twol\ corrections
to the $\MW$--$\MZ$~interdependence~\cite{dr2lB}. Contrary to the SM case, 
these \twol\ strong corrections turned out to increase the \onel\
contributions, leading to an enhancement of up to 35\%~\cite{dr2lA}.
Most recently, the leading \twol\ contributions to 
$\De\rho$ at \order{\al_t^2}, \order{\al_t \al_b}, \order{\al_b^2},
i.e.\ the leading two-loop contributions involving the top and bottom
Yukawa couplings, have been evaluated~\cite{dr2lC}. They affect
$\MW$ and $\sweff$ by shifts reaching  $12 \mev$ and 
$5 \times 10^{-5}$, respectively. 

At the quantum level, the Higgs sector of the MSSM is considerably
affected by loop contributions and makes $m_h$ yet another sensitive observable.
Precise predictions for the mass $m_h$ of the lightest Higgs boson  
$h$ and its couplings to other particles in terms
of the relevant SUSY parameters are necessary in order to determine the
discovery and exclusion potential of the upgraded Tevatron, and
for physics at the LHC and a future linear collider, where 
high-precision measurements of the Higgs-boson(s) profile
will become feasible~\cite{tesla,nlc,jlc}.

Radiative corrections to the Higgs-boson masses in the $\cp$-conserving
MSSM with minimal flavor violation (MFV) are meanwhile quite
advanced. Besides the full \onel\ 
corrections~\cite{mhiggsf1lB,mhiggsf1lC}, the \twol\ corrections have
been evaluated in the effective-potential 
method~\cite{mhiggsEP1,mhiggsEPFD,mhiggsEP2,mhiggsEP3}, the
renormalization-group 
approach~\cite{mhiggsRG}, and the Feynman-dia\-gram\-matic
approach~\cite{mhiggsletter,mhiggslong,mhiggslle} (see
\cite{bse,mhiggsRGFD2} for a comparison), providing all
leading \twol\ contributions available by now~\cite{mhiggsAEC}. 
However, the impact of non-minimal flavor violation (NMFV) on the
MSSM Higgs-boson masses and mixing angles, entering already at the \onel\
level, has not been explored so far, although 
effects from possible NMFV on Higgs-boson decays
were investigated in \cite{HdecNMFV,HdecNMFV2}. Simultaneously, 
effects of NMFV enter also the electroweak precision observables
at the \onel\ level, but have never been analyzed as yet. 
Hence, we study in this paper the consequences from NMFV 
for both the electroweak precision observables and the MSSM
lightest Higgs-boson mass $m_h$.

The most general flavor structure of the soft SUSY-breaking
sector with flavor non-diagonal terms would
induce large flavor-changing neutral-currents, contradicting
the experimental results~\cite{pdg}. Attempts to avoid this kind of
problem include flavor-blind SUSY-breaking scenarios, like minimal
Supergravity or gauge-mediated SUSY-breaking. In these
scenarios, the sfermion-mass matrices are flavor diagonal in the same
basis as the quark matrices at the SUSY-breaking scale. However, a certain
amount of flavor mixing is generated due to the renormalization-group
evolution from the SUSY-breaking scale down to the electroweak
scale. Estimates of this radiatively induced off-diagonal squark-mass
terms indicate that the largest entries are those connected to the
SUSY partners of the left-handed quarks~\cite{NMFVestimate,savoy},
generically denoted as $\De_{LL}$. Those off-diagonal soft
SUSY-breaking terms scale with the square of diagonal soft
SUSY-breaking masses $\msusy$, whereas the
$\De_{LR}$ and $\De_{RL}$ terms scale linearly, and $\De_{RR}$ with
zero power of $\msusy$. Therefore, usually the hierarchy 
$\De_{LL} \gg \De_{LR,RL} \gg \De_{RR}$ is realized. It was also
shown in \cite{NMFVestimate,savoy} that mixing between the third and
second generation squarks can be numerically significant due to the
involved third-generation Yukawa couplings.
On the other hand, there are strong experimental bounds on squark
mixing involving the first generation, coming from data on 
$K^0$--$\bar K^0$ and 
$D^0$--$\bar D^0$~mixing~\cite{FirstGenMix,FirstGenMix2}.

The analytical results obtained in this paper have been derived for
the general case of mixing between
the third and second generation of squarks,
i.e.\ all NMFV contributions, $\De_{LL,LR,RL,RR}$, can be chosen
independently in the $\Stop/\Scha$ and in the $\Sbot/\Sstr$ sector 
(corrections from the first-generation squarks are not considered, for
reasons mentioned above). The numerical analysis of NMFV effects, however, 
and the illustration of the behavior of $m_h$ and electroweak observables
are performed for the simpler, but well motivated, scenario (also 
chosen in~\cite{HdecNMFV2}) where only mixing between 
$\StopL$ and $\SchaL$ as well as between $\SbotL$ and $\SstrL$ is considered,
with $\De_{LL}^{t}$ and $\De_{LL}^{b}$ as
the only flavor off-diagonal entries in the squark-mass matrices.

\bigskip
The paper is organized as follows.
In~\refse{sec:NMFV} we review the MSSM with NMFV
and set up the notation. Corrections to the lightest MSSM Higgs-boson mass 
at the \onel\ level arising from NMFV are presented 
in~\refse{sec:mh}. Analytical and numerical results
for $\De\rho$ are given in~\refse{sec:ewpo}, together with a numerical 
analysis of the full one-loop effects from scalar quarks on $\MW$ and
$\sweff$. \refse{sec:conclu} is devoted to the conclusions. 
Finally, in the appendix, we list the set of Feynman rules 
for the general case of NMFV.

\section{Non-minimal flavor violation in the MSSM}
\label{sec:NMFV}

As explained in the introduction, our analytical results are obtained
for a general mixing of the third and second generation of scalar
quarks. The squark mass matrices in the basis of 
$(\SchaL, \StopL, \SchaR, \StopR)$ and 
$(\SstrL, \SbotL, \SstrR, \SbotR)$~\footnote{
Note that our convention is slightly different from the one 
used in \cite{HdecNMFV2}.}%
~are given by
\BEA
\label{eq:massup}
M_{\tiu}^2 &=& 
\MLv 
M_{\tilde L_c}^2 & \De_{LL}^t & \mc \Xc  & \De_{LR}^t \\[.3em]
\De_{LL}^t & M_{\tilde L_t}^2 & \De_{RL}^t &\mt \Xt \\[.5em]
\mc \Xc & \De_{RL}^t & M_{\tilde R_c}^2 & \De_{RR}^t \\[.3em]
\De_{LR}^t & \mt \Xt & \De_{RR}^t & M_{\tilde R_t}^2 
\MR \\[1em]
\label{eq:massdown}
M_{\tid}^2 &=& 
\MLv 
M_{\tilde L_s}^2 & \De_{LL}^b & \ms \Xs &  \De_{LR}^b \\[.3em]
\De_{LL}^b & M_{\tilde L_b}^2 & \De_{RL}^b & \mb \Xb  \\[.5em]
\ms \Xs & \De_{RL}^b & M_{\tilde R_s}^2 & \De_{RR}^b \\[.3em]
\De_{LR}^b & \mb \Xb & \De_{RR}^b & M_{\tilde R_b}^2 
\MR
\EEA
with
\BEA
\label{eq:defMXt}
M_{\tilde L_q}^2 &=& M_{\tilde Q_q}^2 + \mq^2 + 
                     \CZb \MZ^2 (T_3^q - Q_q \sw^2) \nn \\
M_{\tilde R_q}^2 &=& M_{\tilde U_q}^2 + \mq^2 + 
                     \CZb \MZ^2 Q_q \sw^2 ~(q = t,c) \nn \\
M_{\tilde R_q}^2 &=& M_{\tilde D_q}^2 + \mq^2 + 
                     \CZb \MZ^2 Q_q \sw^2 ~(q = b,s) \nn \\
X_q &=& \Aq - \mu (\tb)^{-2 T_3^q}
\EEA
where $m_q$, $Q_q$ and $T_3^q$ are the mass, electric charge and
weak isospin of the quark~$q$. 
$M_{\tilde Q_q}$, $M_{\tilde U_q}$, $M_{\tilde D_q}$  are the soft
SUSY-breaking parameters. The $SU(2)$~structure of the model requires 
$M_{\tilde Q_q}$ to be equal for $\Stop$ and $\Sbot$ as well
as for $\Scha$ and $\Sstr$. 
The expressions furthermore contain the $Z$ and $W$ boson masses $M_{Z,W}$;
the electroweak mixing angle in $s_W = \sin\theta_W$, $c_W = \cos\theta_W$;
the trilinear Higgs couplings $A_q\, (q = t, b, c, s)$ 
to $\Stop$, $\Sbot$, $\Scha$, $\Sstr$; the Higgsino mass  parameter $\mu$, 
and  $\tb = v_2/v_1$.

In order to diagonalize the two $4 \times 4$~squark mass matrices, two 
$4 \times 4$~rotation matrices, $R_{\tiu}$ and $R_{\tid}$, are needed,
\BE
\tiu_{\al} \; = \; R_{\tiu}^{\al,j} \VL \SchaL \\\StopL \\
                                      \SchaR \\ \StopR \VR_j ~,~~~~
\tid_{\al} \; = \; R_{\tid}^{\al,j} \VL \SstrL \\ \SbotL \\
                                      \SstrR \\ \SbotR \VR_j ~,
\label{newsquarks}
\end{equation}
yielding the diagonal mass-squared matrices as follows,
\BEA
{\rm diag}\{m_{\tiu_1}^2, m_{\tiu_2}^2, 
          m_{\tiu_3}^2, m_{\tiu_4}^2 \}^{\al,\be} & = &
R_{\tiu}^{\al,i} \; \KL M_{\tiu}^2 \KR_{i,j} \; 
( R_{\tiu}^{\be,j} )^\dagger ~,\\
{\rm diag}\{m_{\tid_1}^2, m_{\tid_2}^2, 
          m_{\tid_3}^2, m_{\tid_4}^2 \}^{\al,\be} & = &
R_{\tid}^{\al,i} \; \KL M_{\tid}^2 \KR_{i,j} \; 
( R_{\tid}^{\be,j} )^\dagger ~.
\EEA

Feynman rules that involve two scalar quarks can be obtained from the
rules given in the $\sfl,\sfr$~basis by applying the corresponding
rotation matrix ($\tiq =\tiu,\tid$), 
\BE
V(X\tiq_{\al}\tiq_{\be}^\prime) \; = \; 
 R_{\tiq}^{\al,i} \;  R_{\tiq^\prime}^{\be,j} \; 
 V(X\tiq_i \tiq^\prime_j)~.
\end{equation}
Thereby $V(X\tiq_i \tiq^\prime_j)$ denotes a generic vertex in the 
$\sfl,\sfr$~basis, and $V(X\tiq _{\al}\tiq_{\be}^\prime)$ is the
vertex in the NMFV mass-eigenstate basis.
The Feynman rules for the vertices
needed  for our applications, i.e.\ the interaction of one
and two Higgs or gauge bosons with two squarks, can be found in the
appendix. This new set of generalized vertices has been 
implemented into the program packages
\fa/{\em FormCalc}\cite{feynarts} extending the previous MSSM model
file~\cite{famssm}~\footnote{
The model file is available on request.}. The extended \fa\ version  
was used for the evaluation of the Feynman diagrams along this
paper to obtain the general analytical results.

For the numerical analysis we are more specific and consider the 
simpler scenario with mixing only between the left-handed components
of $\Stop,\Scha$  and  $\Sbot,\Sstr$, as explained in the introduction.
The only flavor off-diagonal entries 
in the squark-mass matrices are normalized according to  
$\De_{LL}^{t,b} = \la^{t,b} M_{\tilde Q_3} M_{\tilde Q_2}$,
following~\cite{NMFVestimate,FirstGenMix,FirstGenMix2}~\footnote
    {The parameters $\la^t$ and
    $\la^b$ introduced here are denoted by 
    $(\delta_{{\scriptsize{LL}}}^u)_{23}$ and 
    $(\delta_{LL}^d)_{23}$ 
    in~\cite{NMFVestimate,FirstGenMix,FirstGenMix2}. 
}, where $M_{\tilde Q_3, \tilde Q_2}$ are the soft SUSY-breaking
masses for the $SU(2)$ squark doublet in the third and second
generation. NMFV is thus parametrized in terms of the dimensionless
quantities $\la^t$ and $\la^b$ (see
\cite{FirstGenMix,FirstGenMix2,expconst2,expconst3} for experimentally
allowed ranges).
The case of $\la^t = \la^b = 0$ corresponds to the MSSM with
minimal flavor violation (MFV). In detail, we have
\begin{align}
\De_{LL}^t &= \la^t M_{\tilde L_t} M_{\tilde L_c} \,, &
\De_{LR}^t = \De_{RL}^t = \De_{RR}^t &= 0 \,, \nn \\
\De_{LL}^b &= \la^b M_{\tilde L_b} M_{\tilde L_s} \,, &
\De_{LR}^b = \De_{RL}^b = \De_{RR}^b &= 0 \, ,
\end{align}
for the entries in the matrices (\ref{eq:massup}) and (\ref{eq:massdown}).

For the sake of simplicity, we have assumed in our numerical analysis
the same flavor mixing parameter in the $\Stop - \Scha$ and 
$\Sbot - \Sstr$ sectors, $\la = \la^t = \la^b$. 
It should be noted in this respect that LL
blocks of the up-squark and down-squark mass matrices are not
independent because of the $SU(2)$ gauge invariance; they are related
trough the CKM mass matrix~\cite{FirstGenMix2}, which also implies that
a large difference between these two parameters is not allowed.

\section{The mass of the lightest Higgs boson}
\label{sec:mh}

The higher-order corrected masses $m_h,\, m_H$ of the $\cp$-even 
neutral Higgs bosons $h,H$ correspond to the poles of the $h,H$-propagator 
matrix. In terms of its inverse, it is given by
\BE
\left(\De_{\rm Higgs}\right)^{-1}
= - i \ML p^2 -  \mHtree^2 + \hSi_{HH}(p^2) &  \hSi_{hH}(p^2) \\[.5em]
     \hSi_{hH}(p^2) & p^2 -  \mhtree^2 + \hSi_{hh}(p^2) \MR,
\label{higgsmassmatrixnondiag}
\end{equation}
where $\mhtree, \mHtree$ are the tree-level $h,H$ masses,
and $\hSi(p^2)$ denote the renormalized
Higgs-boson self-energies for a general momentum $p$.
Determining the poles of the matrix $\De_{\rm Higgs}$ in
(\ref{higgsmassmatrixnondiag}) is equivalent to solving
the equation
\begin{equation}
\left[p^2 - \mhtree^2 + \hSi_{hh}(p^2) \right]
\left[p^2 - \mHtree^2 + \hSi_{HH}(p^2) \right] -
\left[\hSi_{hH}(p^2)\right]^2 = 0\,.
\label{eq:proppole}
\end{equation}
The status of the available results for the self-energy contributions to
(\ref{higgsmassmatrixnondiag}) has been summarized in the introduction
(see also \cite{mhiggsAEC} for a review). 

\begin{figure}[t!]
\begin{center}
\includegraphics[width=13.5cm,clip=]{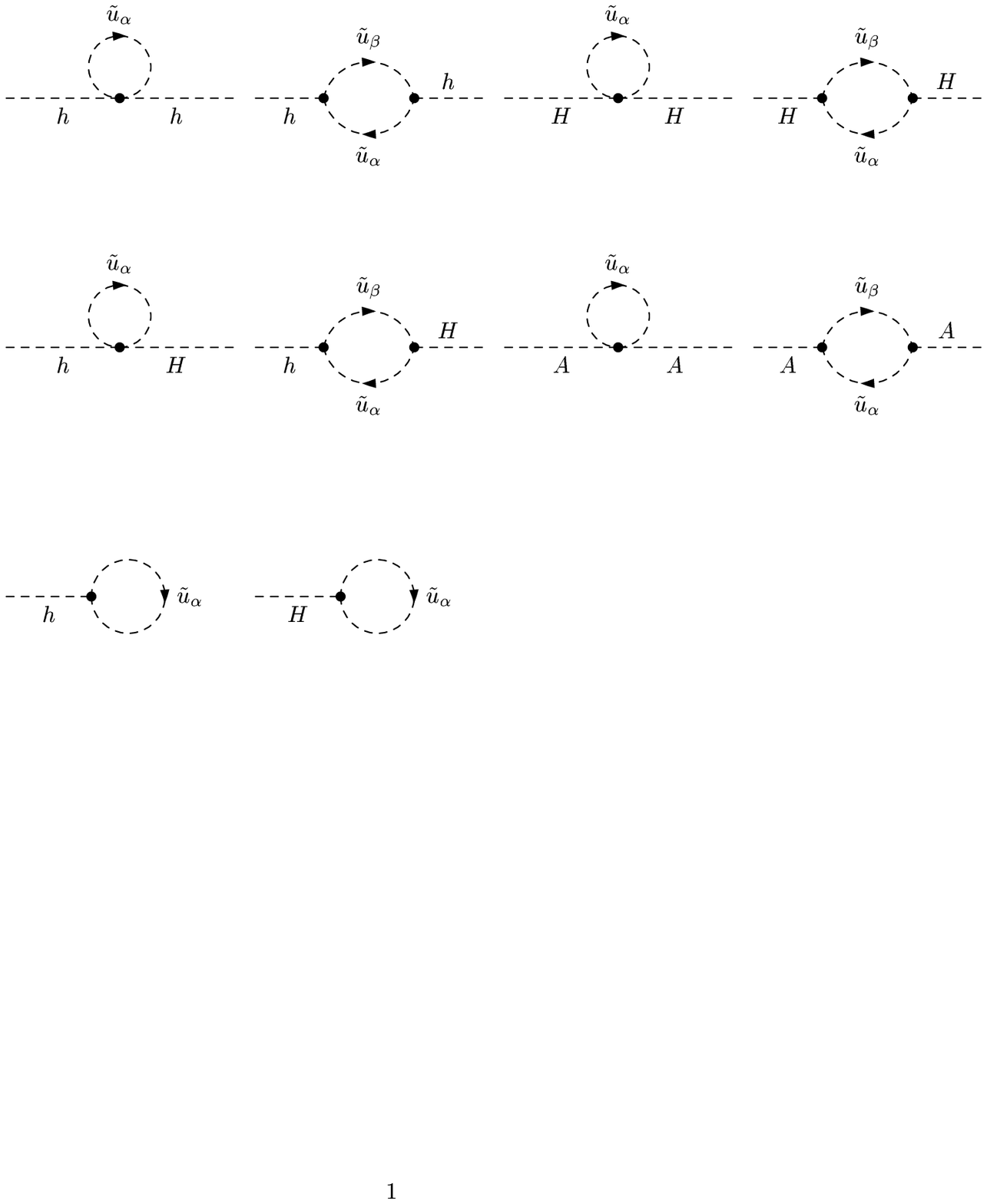}
\end{center}\vspace*{-1cm}
\caption[]{
Feynman diagrams for the squark contributions to the 
Higgs boson self-energies and for the tadpole contributions.
}
\label{fig:FDse}
\end{figure}
%
Within the MSSM with MFV, the dominant \onel\ contributions to the 
self-energies in~(\ref{higgsmassmatrixnondiag})
result from the Yukawa part of the theory (i.e.\
neglecting the gauge couplings); 
they are described  by loop diagrams involving 
third-generation quarks and squarks.
Within the MSSM with NMFV, the squark loops have to be modified
by introducing the generation-mixed squarks, as given in (\ref{newsquarks}).
The contributing Feynman diagrams are illustrated in~\reffi{fig:FDse}.
The leading terms are obtained by evaluating the contributions to 
the renormalized Higgs-boson self-energies at zero external momentum, 
$\hSi_s(0), s = hh, hH, HH$. 
Thereby, the renormalized self-energies are given by
\BE
\hSi_s = \Si_s - \de V_s ~,~~s = hh\,, hH\,, HH\,.
\end{equation}
$\Si_s$ are the unrenormalized Higgs boson self-energies, and
$\de V_s$ are the counter terms for the various coefficients in 
the quadratic part of the Higgs potential,
\BEA
\label{hhren}
\de V_{hh} &=& \de\MA^2 (\ca\cbe + \sa\sbe)^2 \non\\
 && - T_1 \frac{e}{2\sw\MW} 
      (-2 \ca \sa \sbe^3 + \cbe (-\ca^2\sbe^2 + \sa^2 (1 + \sbe^2))) \non\\
 && - T_2 \frac{e}{2\sw\MW} 
      (-2 \ca \sa \cbe^3 + \sbe (\ca^2 (1 + \cbe^2) - \sa^2\cbe^2))\,, \nn\\
\label{HHren}
\de V_{HH} &=& \de\MA^2 (\sa\cbe - \ca\sbe)^2 \non \\
 && - T_1 \frac{e}{2\sw\MW}
      (-\cbe \sa^2 \sbe^2 + 2 \sa\ca\sbe^3 + \ca^2 \cbe (1 + \sbe^2)) \non\\
 && - T_2 \frac{e}{2\sw\MW}
      (2 \sa\ca\cbe^3 - \ca^2\cbe^2\sbe + (1 + \cbe^2) \sa^2\sbe)\,, \nn\\
\label{hHren}
\de V_{hH} &=& \de\MA^2 (\sbe\cbe (\sa^2 - \ca^2)
                                   + \sa\ca (\cbe^2 - \sbe^2)) \non \\
 && - T_1 \frac{e}{2\sw\MW} 
      (\sbe^3 (\ca^2 - \sa^2) - \sa\ca\cbe (1 + 2 \sbe^2)) \non \\
 && - T_2 \frac{e}{2\sw\MW}
      (\cbe^3 (\ca^2 - \sa^2) + \sa\ca\sbe (1 + 2 \cbe^2))\,. 
\EEA

These expressions involve $\sa \equiv \Sa$, $\ca \equiv \Ca$
of the angle $\alpha$ diagonalizing the lowest-order Higgs-boson
mass matrix, the $A$-boson mass counter term, and the tadpoles 
$T_1$ and $T_2$. In the on-shell renormalization scheme (in the 
leading Yukawa approximation) they are determined by
\begin{equation}
\de\MA^2 = \Si_A(0)
\end{equation}
\BE
\hspace*{-5cm}
{\mbox{and}}\hspace*{5cm}
T_1 = T_{H \big| \al \to 0}~, \qquad T_2 = T_{h \big| \al \to 0} \, ,
\end{equation}
where $T_{h,H}$ correspond to the tadpole diagrams displayed 
in~\reffi{fig:FDse}.

Here we restrict ourselves to the dominant Yukawa
contributions resulting from the top and $t/\Stop$ (and $c/\Scha$) sector.
Corrections from $b$ and $b/\Sbot$ (and
$s/\Sstr$) could only be important for very large values of
$\tb$, $\tb \gsim \mt/\mb$, which we do not consider here.
The analytical result of the renormalized Higgs boson self-energies,
based on the general $4 \times 4$~structure of the $\Stop/\Scha$~mass
matrix, has then been
implemented into the Fortran code \fhto~\cite{feynhiggs} that
includes all existing higher-order corrections (of the MFV MSSM). All
data shown in this letter has then been obtained with the help of \fhto.

The results for the lightest MSSM Higgs-boson mass, including all
available corrections also at the \twol\ level, are presented for
five benchmark scenarios defined in \cite{LHbenchmark}, 
named ``$\mhmax$'' (to maximize the lightest Higgs boson mass), 
``constrained $\mhmax$'' (labeled as ``$\Xt/\msusy = -2$''),
``no-mixing'' (with no mixing in the MFV $\tilde t$ sector), 
``gluophobic Higgs'' (with reduced $ggh$ coupling), 
and ``small~$\aeff$'' scenario (with reduced $h b \bar b$ and $h \tau^+
\tau^-$ couplings).
For all these benchmark scenarios the soft SUSY-breaking parameters in
the three generations of scalar quarks are equal,
\BE
\msusy = M_{\tilde Q_q} = M_{\tilde U_q} =M_{\tilde D_q}~,
\end{equation}
as well as  all the trilinear couplings, $A_s=A_b=A_c=A_t$. 
Despite these simplifications, the
five scenarios can show quite different behavior concerning
observables in the Higgs sector~\cite{LHbenchmark}. 

\begin{figure}[tb!]
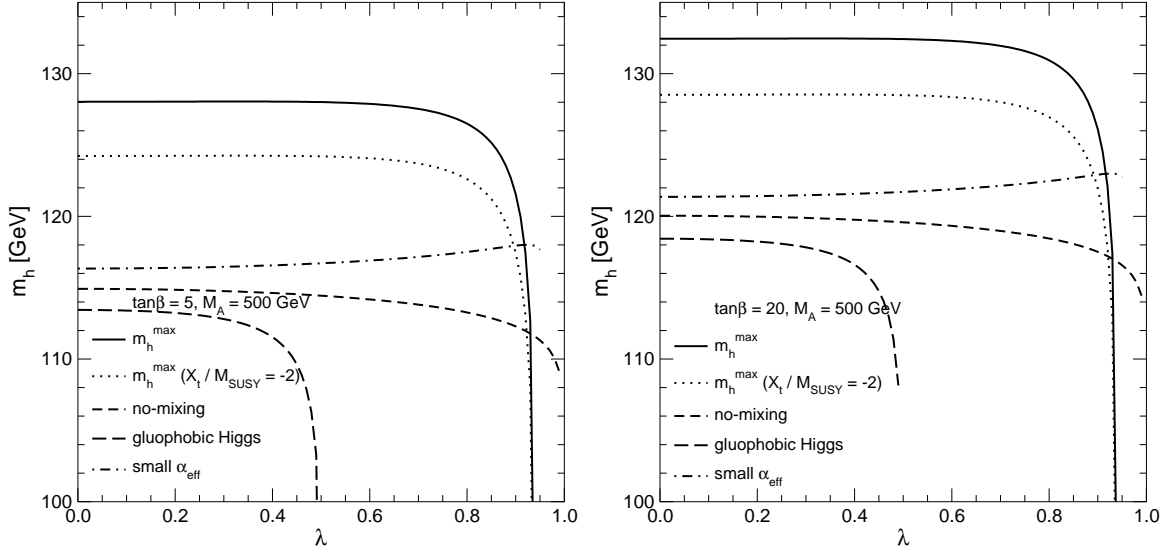

\begin{center}
\epsfig{figure=mhlam03.bw.eps,width=7.6cm}
\epsfig{figure=mhlam01.bw.eps,width=7.6cm}
\end{center}\vspace*{-0.4cm}
\caption[]{
The variation of $\mh$ with $\la = \la^t$ is shown in five
benchmark scenarios~\cite{LHbenchmark}.
$\MA$ has been fixed to $\MA = 500 \gev$, and $\tb$ is set to 
$\tb = 5$ (left panel) or $\tb = 20$ (right panel). }
\label{fig:mhmax}
\end{figure}

In \reffi{fig:mhmax} we illustrate the dependence of
$\mh$ on $\la( = \la^t)$ in all five benchmark scenarios.
$\MA$ has been fixed to $\MA = 500 \gev$, and $\tb$ is set to 
$\tb = 5$ (left) or $\tb = 20$ (right). 
All scenarios show a similar behavior. For small to moderate allowed
values of~$\la$ the variation of $\mh$ is small. Only for large values
(around 0.5 in the gluophobic Higgs scenario, and around 0.9 in
the other four scenarios) the variation of $\mh$ can be quite strong,
up to the \order{5 \gev}. In the gluophobic Higgs scenario unphysical
values for the scalar quark masses are reached already for smaller
values of~$\la$, since $\msusy$ is quite low in this scenario (see
\cite{LHbenchmark} for details). 
Values of $\lambda$ above $0.5$ imply forbidden values for the
squark masses in this scenario. In all cases except for the
small~$\aeff$ scenario the lightest Higgs boson mass turns out to be
reduced. In the small~$\aeff$ scenario it can be enhanced by up to 
$2 \gev$. Considering that large values of $\la$ are in conflict 
with FCNC data, the impact of NMFV on $\mh$ is in general rather small.
Conversely, independent of low-energy FCNC data on flavor mixing, 
high values of $\la$ can be constrained by 
the experimental lower bound on $\mh$~\cite{LEPHiggs}.

\section{\boldmath{$\De\rho$} and electroweak precision observables}
\label{sec:ewpo}

\psfrag{lam}{{\scriptsize $\la$}}
\psfrag{lamu}{{\scriptsize $\la _{\tiu}$}}
\psfrag{lamd}{{\scriptsize $\la _{\tid}$}}
\psfrag{lamt}{{\scriptsize $\la_{t}$}}
\psfrag{lamb}{{\scriptsize $\la_{b}$}}
\psfrag{dr}{{\scriptsize $\De \rho $}}
\psfrag{MSusy}{{\scriptsize $M_{\rm {\tiny{SUSY}}} $ [GeV]}}
\psfrag{mhmax}{{\scriptsize $m_{h}^{\mbox{\tiny{max}}}$ scenario}}
\psfrag{nomix}{{\scriptsize no-mixing scenario }}
\psfrag{mhmax1000}{{\scriptsize $m_{h}^{\mbox{\tiny{max}}},
M_{\mbox{\fontsize{1}{1}\selectfont SUSY \normalfont}}=1 \tev $}}
\psfrag{nomix1000}{{\scriptsize no-mixing,
$M_{\mbox{\fontsize{1}{1}\selectfont SUSY \normalfont}}=1 \tev $}}
\psfrag{mhmax2000}{{\scriptsize $m_{h}^{\mbox{\tiny{max}}},
M_{\mbox{\fontsize{1}{1}\selectfont SUSY \normalfont}}=2 \tev $}}
\psfrag{nomix2000}{{\scriptsize no-mixing,
$M_{\mbox{\fontsize{1}{1}\selectfont SUSY \normalfont}}=2 \tev $}}
\psfrag{mhmaxep}{{\scriptsize $m_{h}^{\mbox{\tiny{max}}},\,$ \eps
    positive}}
\psfrag{mhmaxem}{{\scriptsize $m_{h}^{\mbox{\tiny{max}}},\,$ \eps
    negative}}
\psfrag{nomixep}{{\scriptsize no-mixing,\,\eps positive}}
\psfrag{nomixem}{{\scriptsize no-mixing, \,\eps negative}}
\psfrag{nomix}{{\scriptsize no-mixing scenario }}
\psfrag{dmw}{\rotatebox{90}{\footnotesize $\de \MW \, [GeV]$}}
\psfrag{dst}{\rotatebox{90}{\hspace*{0.5cm}\footnotesize $\de \sweff$  }}
\psfrag{f1}{$f_1$}
\psfrag{f2}{$f_2$}

One important consequence of flavor mixing through the 
flavor non-diagonal entries in the squark mass 
matrices~(\ref{eq:massup},\ref{eq:massdown}) is to generate large splittings 
between the squark-mass eigenvalues. The loop contribution to the 
electroweak $\rho$ parameter,
\BE
\De\rho = \frac{\Si_Z(0)}{\MZ^2} - \frac{\Si_W(0)}{\MW^2} ,
\label{delrho}
\end{equation}
with the unrenormalized $Z$ and $W$ boson self-energies at zero momentum,
$\Si_{Z,W}(0)$, represents the leading universal corrections to the 
electroweak precision observables induced by
mass splitting between partners in isospin doublets~\cite{rho}
and is thus sensitive to the mass-splitting effects induced
by non-minimal flavor mixing. Precisely measured observables~\cite{ewdataw03}
like the $W$ boson mass, $\MW$, and the effective leptonic mixing 
angle, $\sweff$, are affected by shifts according to
\BE
\de\MW \approx \frac{\MW}{2}\frac{\cw^2}{\cw^2 - \sw^2} \De\rho, \quad
\de\sweff \approx - \frac{\cw^2 \sw^2}{\cw^2 - \sw^2} \De\rho .
\label{precobs}
\end{equation}

Within the MSSM with MFV, 
the dominant correction from SUSY particles at the \onel\ level arises from 
the $\Stop$ and $\Sbot$ contributions.
Explicit expressions can be found in~\cite{dr2lC}, together
with the SUSY-QCD and SUSY-EW corrections at two-loop order.

Beyond the $\De\rho$ approximation, the shift in $\MW$ caused 
by a variation of $\De r$ can be written as follows,
\BE
\de\MW = -\frac{\MW}{2} \frac{\sw^2}{\cw^2 - \sw^2} \de(\De r)\,.
\label{delmw}
\end{equation}
As far as $\delta\De r$ originates from loop contributions to the
self energies only, it is given by 
\begin{equation}
\label{deltar}
\de(\De r) = \Si_\ga^\prime(0) 
        - \frac{\cw^2}{\sw^2} \KL \frac{\Si_Z(\MZ^2)}{\MZ^2} -
                                  \frac{\Si_W(\MW^2)}{\MW^2} \KR
        + \frac{\Si_W(0) - \Si_W(\MW^2)}{\MW^2}~,
\end{equation}
with $\Si^\prime =\frac{\partial}{\partial q^2} \Si(q^2)$.
In the case considered here, the self-energies in (\ref{deltar})
stand for the set of squark-loop contributions. Likewise the induced 
shift in the effective mixing angle reads as follows,
\BE
\de\sweff = \frac{\cw^2\,\sw^2}{\cw^2 - \sw^2} \de(\De r) 
            - \sw\cw \KKL \frac{\Si_{\ga Z}(\MZ^2)}{\MZ^2}
                         -\frac{\cw}{\sw} 
                           \KL \frac{\Si_Z(\MZ^2)}{\MZ^2} -
                               \frac{\Si_W(\MW^2)}{\MW^2} \KR \KKR~,
\label{delsw}
\end{equation}
again evaluated for the squark-loop contributions in our case.

\subsection{Analytical results for $\De\rho$}

\begin{figure}[t]
\begin{center}
\includegraphics[width=13.5cm,clip=]{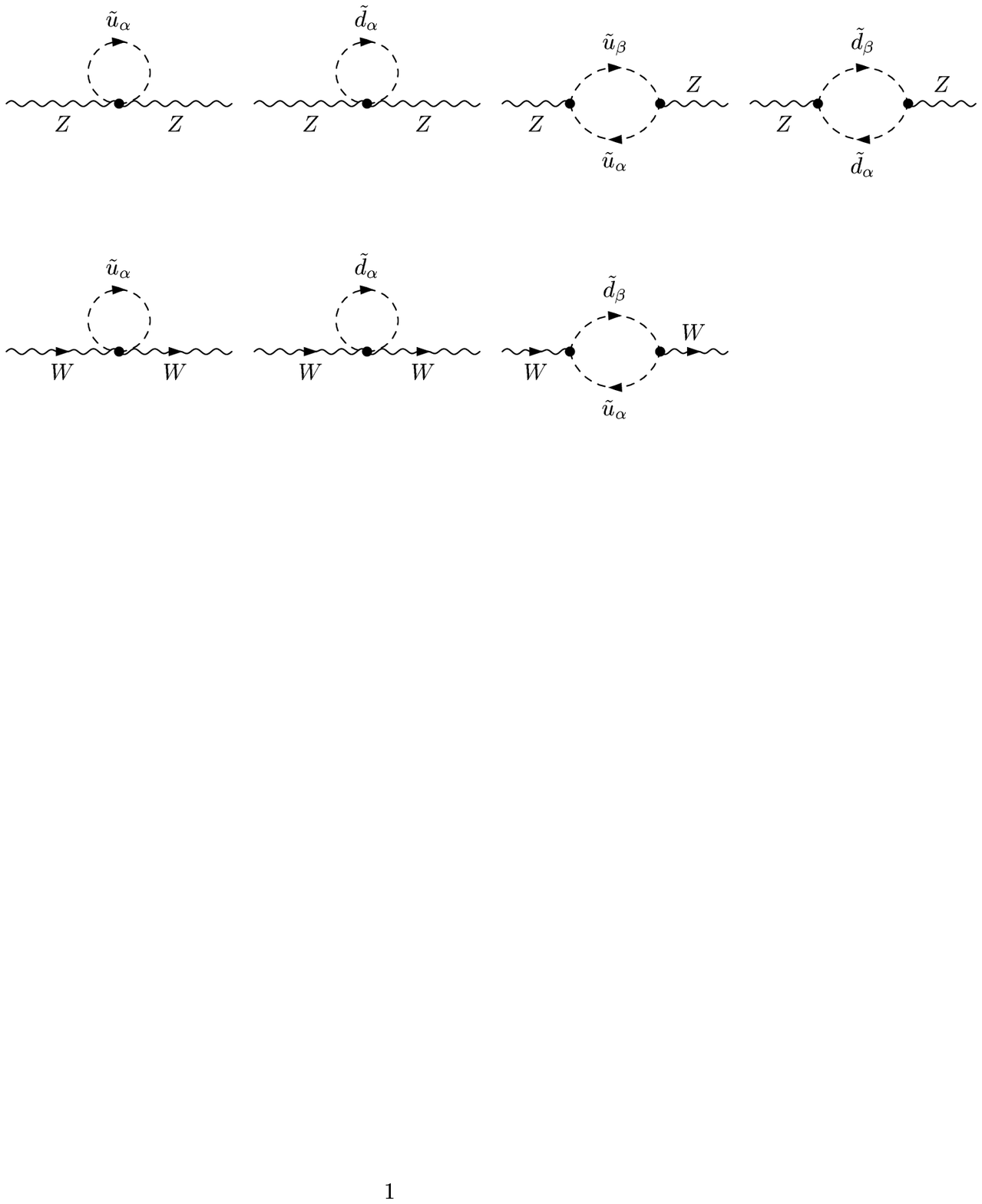}
\end{center}\vspace*{-1cm}
\caption[]{
Feynman diagrams for the squark contributions to the gauge
boson self-energies.}
\label{fig:FDVV}
\end{figure}

Here we consider the supersymmetric NMFV contributions to $\De\rho$ 
resulting from squarks based on the general $4 \times 4$~mass matrix 
for both the $\Stop/\Scha$ and the $\Sbot/\Sstr$ sector, visualized by  
the Feynman diagrams in~\reffi{fig:FDVV}. 
These contributions will be denoted as $\De \rho ^{\tiq}$. 
The analytical \onel\ result for $\De \rho ^{\tiq}$ has been 
implemented into the Fortran code \fhto~\cite{feynhiggs}.

The squark contribution $\De \rho ^{\tiq}$ can be decomposed according to
\BEA
 \De \rho ^{\tiq} = \Xi_Z + \Theta_Z + \Xi_W + \Theta_W \,,
\label{rhosumme}
\EEA
where $\Xi$ and $\Theta$ correspond to different diagram
topologies, i.e.\ to diagrams with trilinear and quartic couplings, 
respectively (see \reffi{fig:FDVV}).
The explicit expressions read as follows,
\BEA
\Xi _W &=& 
 \frac{3 g^2}{8 \pi ^2 \MW^2} \sum \limits_{a,b,c,d} \sum \limits_{\al ,\be} 
 V_{\rm CKM}^{a b} V_{\rm CKM}^{c d} R_{\tiu}^{\al a} R_{\tiu}^{\al c} 
 R_{\tid}^{\be b} R_{\tid}^{\be d} 
 B_{00}(0, m_{\tiu_{\al}}^2, m_{\tid_{\be}}^2) \,,\nn \\
\Theta_W &=& 
  -\frac{3 g^2}{32 \pi ^2 \MW^2} 
  \sum \limits_a \sum \limits_{\al}
   \KKKL (R_{\tiu}^{\al a})^2  A_0(m_{\tiu_{\al}}^2) + 
         (R_{\tid}^{\al a})^2 A_0(m_{\tid_{\al}}^2)\KKKR \,,\nn \\
\Xi_Z &=& 
- \frac{3 g^2}{144 \cwq \pi^2 \MZ^2} \sum \limits_{\al,\be,\ga,\de} 
  \Big\{ \kappa_{\tid}(\ga) R_{\tid}^{\al \ga} R_{\tid}^{\be \ga} 
        \kappa_{\tid}(\de) R_{\tid}^{\al \de} R_{\tid}^{\be \de} 
  B_{00}(0, m_{\tid_{\al}}^2, m_{\tid_{\be}}^2)  \nn \\
&& +    \kappa_{\tiu}(\ga) R_{\tiu}^{\al \ga} R_{\tiu}^{\be \ga} 
        \kappa_{\tiu}(\de) R_{\tiu}^{\al \de} R_{\tiu}^{\be \de} 
  B_{00}(0, m_{\tiu_{\al}}^2, m_{\tiu_{\be}}^2) \Big\} \,,\nn \\
\Theta_Z &=& 
  \frac{3 g^2}{288 \cwq \pi^2 \MZ^2} \sum \limits_{\al,\be,\ga,\de} 
  \KKKL (\kappa_{\tid}(\ga)^2 (R_{\tid}^{\al \ga})^2 
         A_0(m_{\tid_{\al}}^2) + 
         \kappa_{\tiu}(\ga)^2 (R_{\tiu}^{\al \ga})^2 
         A_0(m_{\tiu_{\al}}^2) \KKKR~.
\EEA
Here the indices run from 1 to 2 for Latin letters, and from 1 to 4 for Greek 
letters. The expressions contain the one-point integral $A_0$ and 
the two-point integral $B_{00}$ in $B_{\mu\nu} (k) = g_{\mu\nu}\, B_{00} + 
k_{\mu} k_{\nu} B_{11}$ in the convention of~\cite{feynarts}. 
The remaining constants $\kappa_ {\tiu}$ and $\kappa_ {\tid}$
are defined as follows,
\BEA
\kappa_{\tid} = \VL
3 - 2 \,\swq \\ 3 - 2 \,\swq \\  -2\, \swq \\  -2\, \swq 
\VR~, \qquad \qquad
\kappa_{\tiu} = \VL
-3 + 4 \,\swq \\ -3 + 4 \,\swq \\ 4 \,\swq \\ 4\, \swq
\VR~.
\EEA

The CKM matrix only affects $\Xi_W$. 
Corrections from the first-generation squarks are negligible
due to their very small mass splitting.
Non-minimal flavor mixing of the first generation with the other ones has
been set to zero (see \refse{sec:NMFV}), but conventional CKM mixing
is basically present. Although it is required 
for a UV finite result, it yields only negligibly small effects. Therefore,
for simplification, we drop the first generation and restore the
cancellation of UV divergences by a unitary $2 \times 2$ matrix
replacing the \{23\}-submatrix of the CKM matrix,
\BEA
V_{\rm CKM} = \ML
V_{cs} & V_{cb} \\
V_{ts} & V_{tb} \MR 
= \ML
\cos \epsilon &\sin \epsilon \\
-\sin\epsilon &\cos\epsilon
\MR~,
\label{eq:CKM}
\EEA
with $|\epsilon| \approx 0.04$ close
to the experimental entries~\cite{pdg} of the conventional CKM matrix. 

In the SM (and also in the MSSM with $\la = \la^t = \la^b = 0$) the choice of 
the sign of \eps does not play a role. However, the situation changes
when \lann. In the expression for $\De\rho^{\tiq}$ some terms linear 
in \eps arise from the expansion of $\Xi_W$, and the sign of \eps can
affect the result significantly. The expansion of
$\Xi_W$ can be expressed as,
\begin{equation}
\Xi_W = f_0 + f_1 \epsilon + f_2 \epsilon ^2 + \ldots +
 f_n \epsilon^n + \ldots
\end{equation}
where the coefficients $f_i (i=0,1,2,\ldots)$ are functions of the rotation
matrices $R_{\tilde q}$ and the squarks masses $\msq$ and
therefore, they depend implicitly of the flavor parameter $\la$. 
The explicit analytical expressions for the first terms are:
\BEA
f_0 &=& -\frac{3 g^2}{8 \MW^2 \pi ^2}\sum\limits_{\al, \be} 
 \left( R_{\tid}^{1 \be} R_{\tiu}^{1 \al} +  
        R_{\tid}^{2 \be} R_{\tiu}^{2 \al} \right)^2 
 B_{00}(0, m_{\tiu_\al}^2, m_{\tid_\be}^2) \,,\nn \\
f_1 &=& -\frac{3 g^2}{4 \MW^2 \pi ^2}\sum\limits_{\al, \be} 
 \left( (R_{\tiu}^{1 \al})^2 R_{\tid}^{1 \be}
         R_{\tid}^{2 \be} 
       + R_{\tiu}^{1 \al} R_{\tiu}^{2 \al} 
        (R_{\tid}^{2 \be})^2 \right. \nn \\
&& \left. - R_{\tiu}^{1 \al} R_{\tiu}^{2 \al} 
           (R_{\tid}^{1 \be})^2 -
           (R_{\tiu}^{2 \al})^2 R_{\tid}^{1 \be}
            R_{\tid}^{2 \be}\right) 
 B_{00}(0, m_{\tiu_\al}^2, m_{\tid_\be}^2) \,,\nn \\
f_2 &=& \frac{3 g^2}{8 \MW^2 \pi ^2}\sum\limits_{\al, \be} 
 \left((R_{\tiu}^{1 \al})^2(R_{\tid}^{1 \be})^2 +
       (R_{\tiu}^{1 \al})^2(R_{\tid}^{2 \be})^2 -
       (R_{\tiu}^{2 \al})^2(R_{\tid}^{1 \be})^2 \right. \nn \\
&& \left. - (R_{\tiu}^{2 \al})^2(R_{\tid}^{2 \be})^2 +
             R_{\tiu}^{1 \al} R_{\tiu}^{2 \al} 
             R_{\tid}^{1 \be} R_{\tid}^{2 \be} \right)
 B_{00}(0, m_{\tiu_\al}^2, m_{\tid_\be}^2)\,.
\label{f012}
\EEA

\begin{figure}
\includegraphics[scale=0.8]{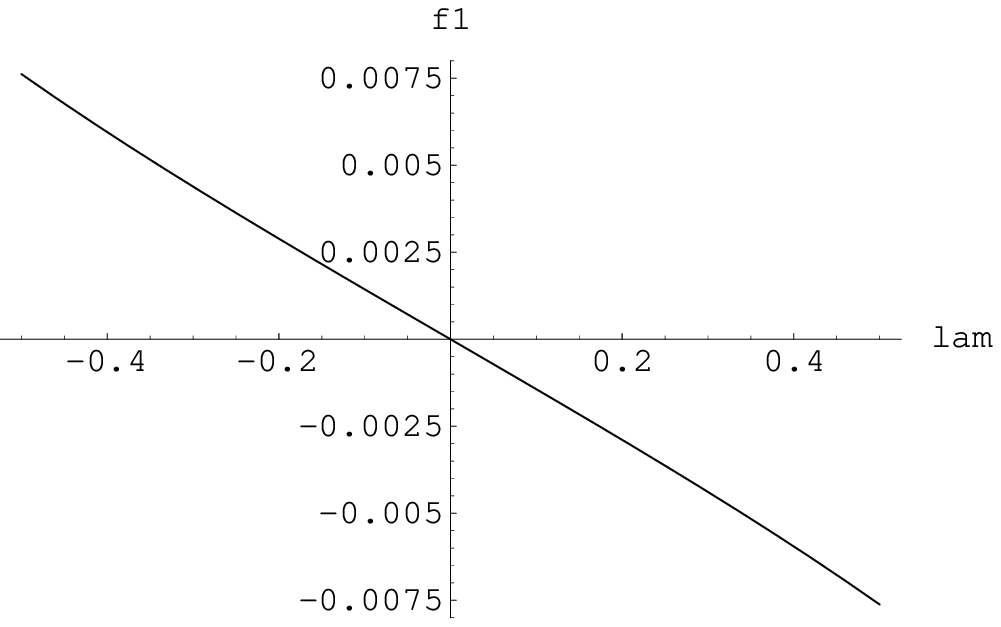}
\includegraphics[scale=0.8]{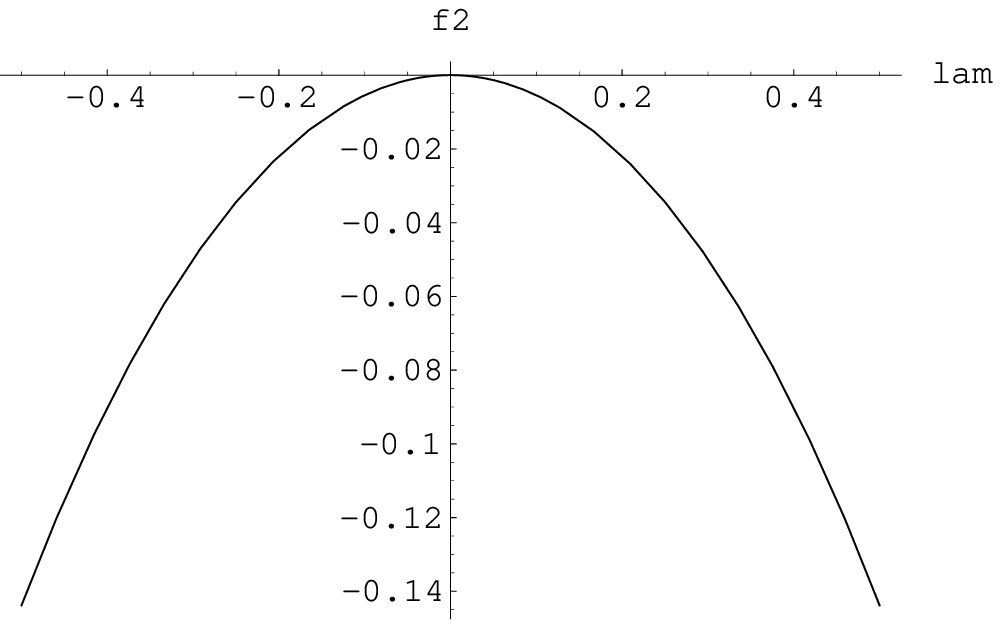}
\caption{$f_1$ and $f_2$ as function of $\la$. SUSY parameters as given
  in~(\ref{eq:inputs}).}
\label{figfs}
\end{figure}

Since $\De\rho^{\tiq}$ is a finite quantity, and the CKM matrix
effects (and therefore, the $\epsilon$ dependence) only appear in $\Xi_W$, 
$f_0$ is the unique coefficient in the expansion that contributes to 
the cancellation of divergences in $\De\rho^{\tiq}$. 
The coefficients $f_1$ and $f_2$ are finite and their $\la$ dependence
is shown in \reffi{figfs}. While $f_1=0$ for $\la =0$, 
$f_2$ is not exactly zero but its value is very small, around 
$5.5 \times 10^{-5}$. This small value at $\la =0$ implies that the CKM effects
in the MSSM with MFV are indeed negligible, which is in agreement with
the universal assumptions in MFV calculations.  
$f_1$ is antisymmetric under $\la \to -\la$, $f_2$ is symmetric, and
so on. Therefore, $\Xi_W$ (and thus $\De\rho$) is symmetric under the
simultaneous reversal of signs $\epsilon \to -\epsilon$, $\la \to -\la$,
i.e.\ only the relative sign has a physical consequence, affecting
the results for $\De\rho$ significantly
(see also \reffi{fig:eps} in the next section). 
In physical terms, non-minimal squark mixing can either strengthen or
partially compensate the CKM mixing.

\subsection{Numerical evaluation of $\De\rho$}
\label{subsec:numevaldeltarho}

For the numerical evaluation, the $\mhmax$ and the no-mixing scenario 
have been selected~\cite{LHbenchmark}, but with a free scale $\msusy$. 
In the $\mhmax$ benchmark scenario the trilinear coupling $\At$ 
is not a free parameter, obeying $\Xt= 2 \msusy$, with $\Xt=\At-\mu\cot\be$.
In the no-mixing scenario, $\At$ is defined by the requirement $\Xt=0$.
The results are independent of $\MA$.
The numerical values of the SUSY parameters are
\BE
\msusy = 1\tev \; {\rm and}\; 2 \tev, \quad \tb = 30, 
\quad \mu=200 \gev, \, \quad \epsilon = 0.04,
\label{eq:inputs}
\end{equation}
if not explicitly stated otherwise. The variation with $\mu$ and $\tb$ 
is very weak, since they do not enter the squark couplings to the vector bosons. 

\begin{figure}[[htb!]
\begin{center}
\vspace*{0.3cm}
\epsfig{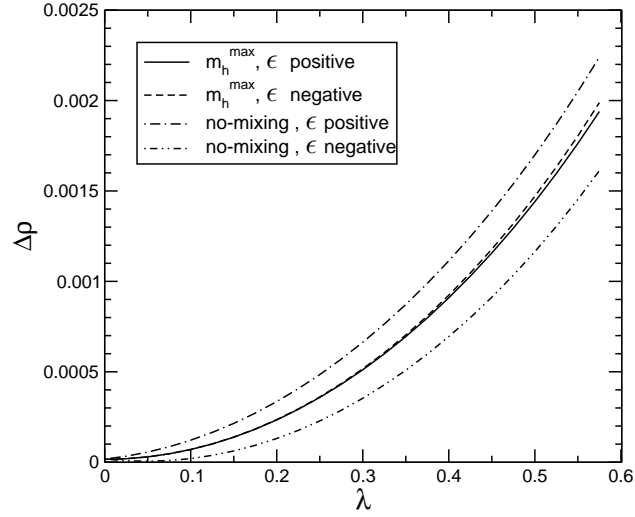}
\end{center}\vspace*{-0.7cm}
\caption[]{
The variation of $\De\rho^{\tiq}$ with $\la (= \la^t = \la^b)$ in the
$\mhmax$ and no-mixing~scenarios for different relative signs of
$\epsilon$ and $\la$. $\msusy = 2 \tev$, 
other SUSY parameters as given in (\ref{eq:inputs}).}
\label{fig:eps}
\end{figure}

To illustrate the above explained 
behavior with the sign of $\epsilon$ explicitly, we show in
\reffi{fig:eps} the corrections to $\De\rho^{\tiq}$ as a function
of $\la (= \la^t = \la^b)$ for different relative signs of \eps and
$\la$, choosing $\la > 0$, and fixing $|\epsilon| =0.04$. 
$\msusy$ has been set to $\msusy = 2 \tev$. For the $\mhmax$~scenario 
the effect is small, but in the no-mixing~scenario the results
are affected significantly by the sign of \eps. 
The squark contribution to $\De\rho^{\tiq}$ can become 
of \order{10^{-3}} for $\la \geq 0.5$.

\psfrag{e}{{\small{$\epsilon$}}}

\begin{figure}[htb!]
\begin{center}
\epsfig{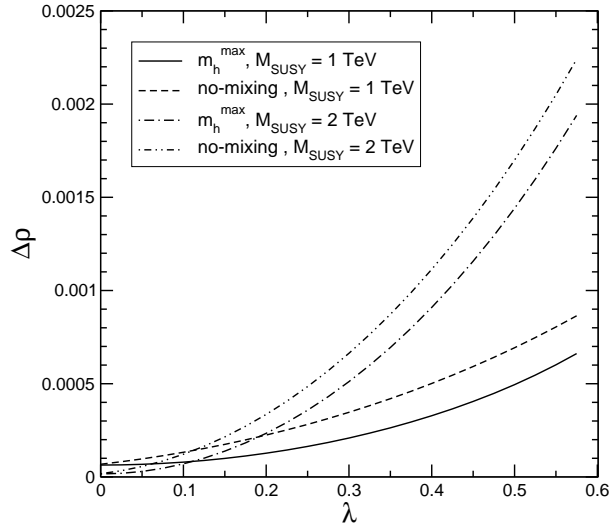}
\end{center}\vspace*{-0.6cm}
\caption[]{
The variation of $\De\rho^{\tiq}$ with $\la = \la^t = \la^b$ in the
$\mhmax$~scenario and no-mixing~scenario. $\msusy$ has been fixed to
$1 \tev$ and $2 \tev$. 
}
\label{fig:rholam}
\end{figure}

\begin{figure}[htb!]
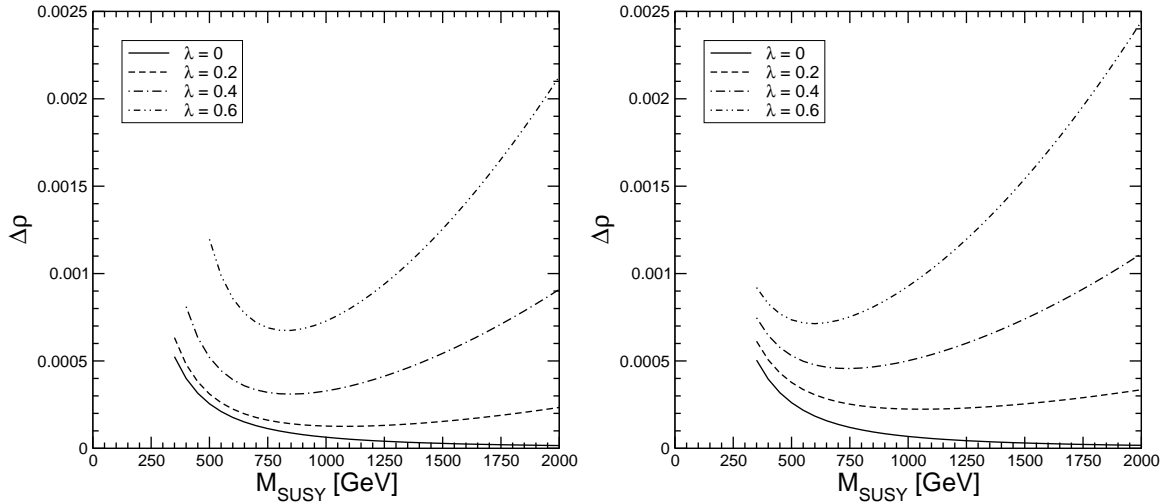

\begin{center}
\vspace*{0.3cm}
\epsfig{figure=drhoMsusy_mhmax.eps,width=7.6cm}
\epsfig{figure=drhoMsusy_nomix.eps,width=7.6cm}
\end{center}\vspace*{-0.6cm}
\caption[]{
The variation of $\De\rho^{\tiq}$ with $\msusy$ 
in the $\mhmax$~scenario (left panel) and
no-mixing~scenario (right panel), for different values of $\la$. }
\label{fig:rhoMSUSY}
\end{figure}

In \reffi{fig:rholam} we show the dependence of $\De\rho^{\tiq}$ on
$\la (= \la^t = \la^b)$ for both the~$\mhmax$ and
no-mixing~scenario and for two values of the SUSY mass scale,
$\msusy=1 \tev$ and $\msusy=2 \tev$. It is clear that
$\De\rho^{\tiq}$ grows with the $\la$ parameter, being close to zero
for $\la = 0$ and $\msusy=2 \tev$. 
One can also see that the effects on $\De\rho^{\tiq}$ are in general
larger for the no-mixing~scenario (see also
the results shown in \citere{dr2lA}).
For large values of $\msusy$ the correction increases with
increasing~$\la$ since the splitting in the squark sector increases.

The behavior of the corrections with the SUSY mass scale is shown 
in \reffi{fig:rhoMSUSY} for different values of $\la$ 
in the $\mhmax$~scenario (left panel) and in the no-mixing~scenario
(right panel). The region below $\msusy \lsim 400 \gev$ (depending on the
scenario) implies too low and hence forbidden values for the
squark masses. The curves are only for the allowed regions.
For $\la = 0$, $\De\rho^{\tiq}$ decreases, being zero
for large $\msusy$ values, in agreement with the results shown in
\citere{dr2lA}. We have also found that, for $\la \neq 0$ and small 
values of $\msusy$, $\De\rho^{\tiq}$ decreases until it reaches 
a minimum and then increases for largest values of the SUSY scale. 
This increasing behavior is more pronounced for larger $\la$ values,
reaching the level of a few per mill.
The reason can be found once again the increasing mass splitting.

We also consider the possibility of choosing different values for $\la^t$
and $\la^b$. We have checked that $\De\rho^{\tiq}$ increases with $\la^t$ 
and $\la^b$ independently, being smallest for $\la^t = \la^b = 0$. 
If $\la^t$ is very different from $\la^b$, 
the values for $\De\rho^{\tiq}$ can be very large. 
For example, for the MSSM parameters we have chosen,
$\De\rho^{\tiq}$ can be as large as $0.08$ for $\la^t=0.6, \la^b=0$.
However, the large splitting between these two parameters is 
disfavored (see the discussion at the end of \refse{sec:NMFV}). 

\subsection{Numerical evaluation for $\MW$ and $\sweff$}
\label{subsec:numevalMWsw2eff}

\begin{figure}[htb!]
\begin{center}
\hspace*{-1cm}
\epsfig{figure=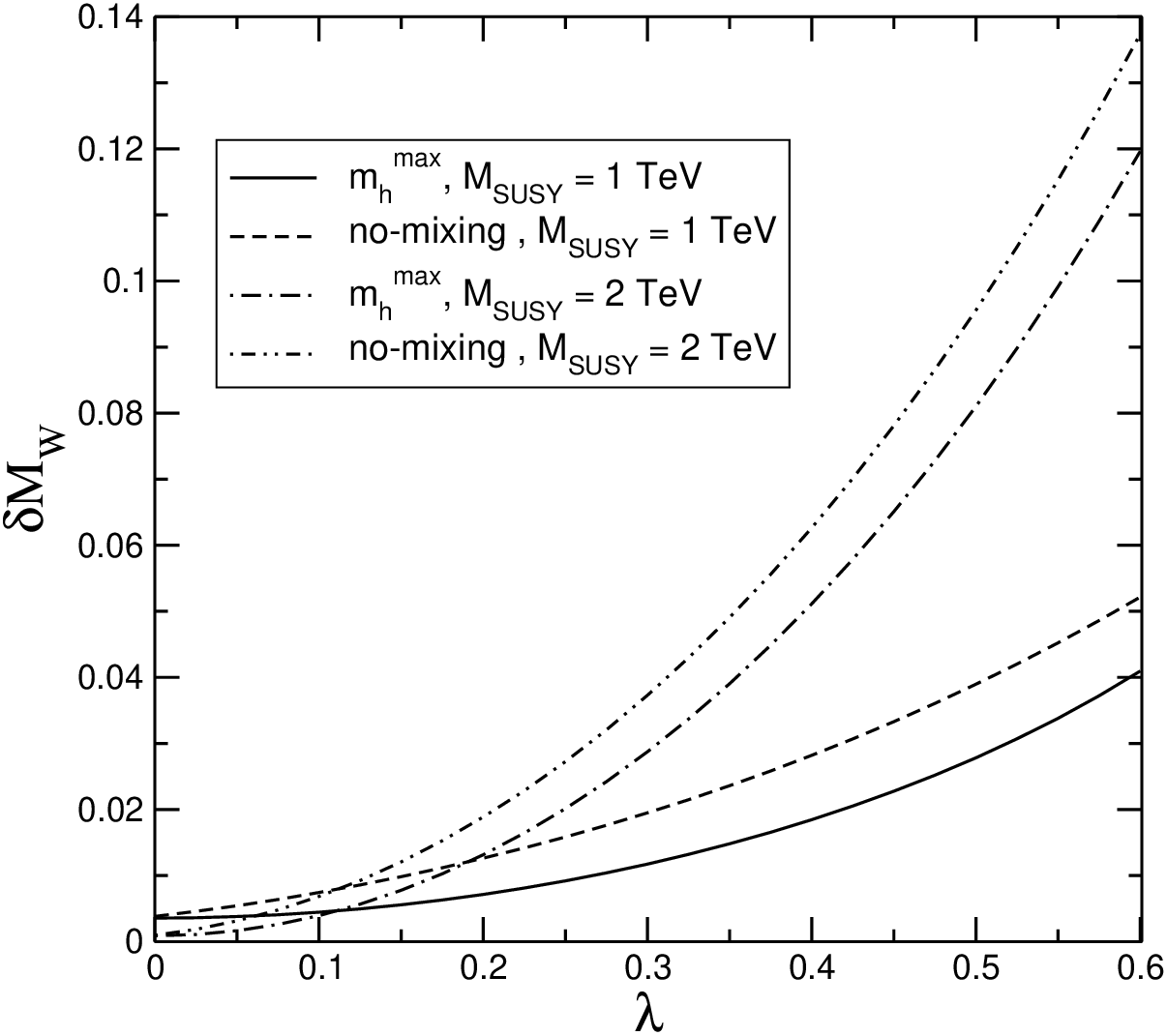,width=7.5cm}\hspace*{0.4cm}
\epsfig{figure=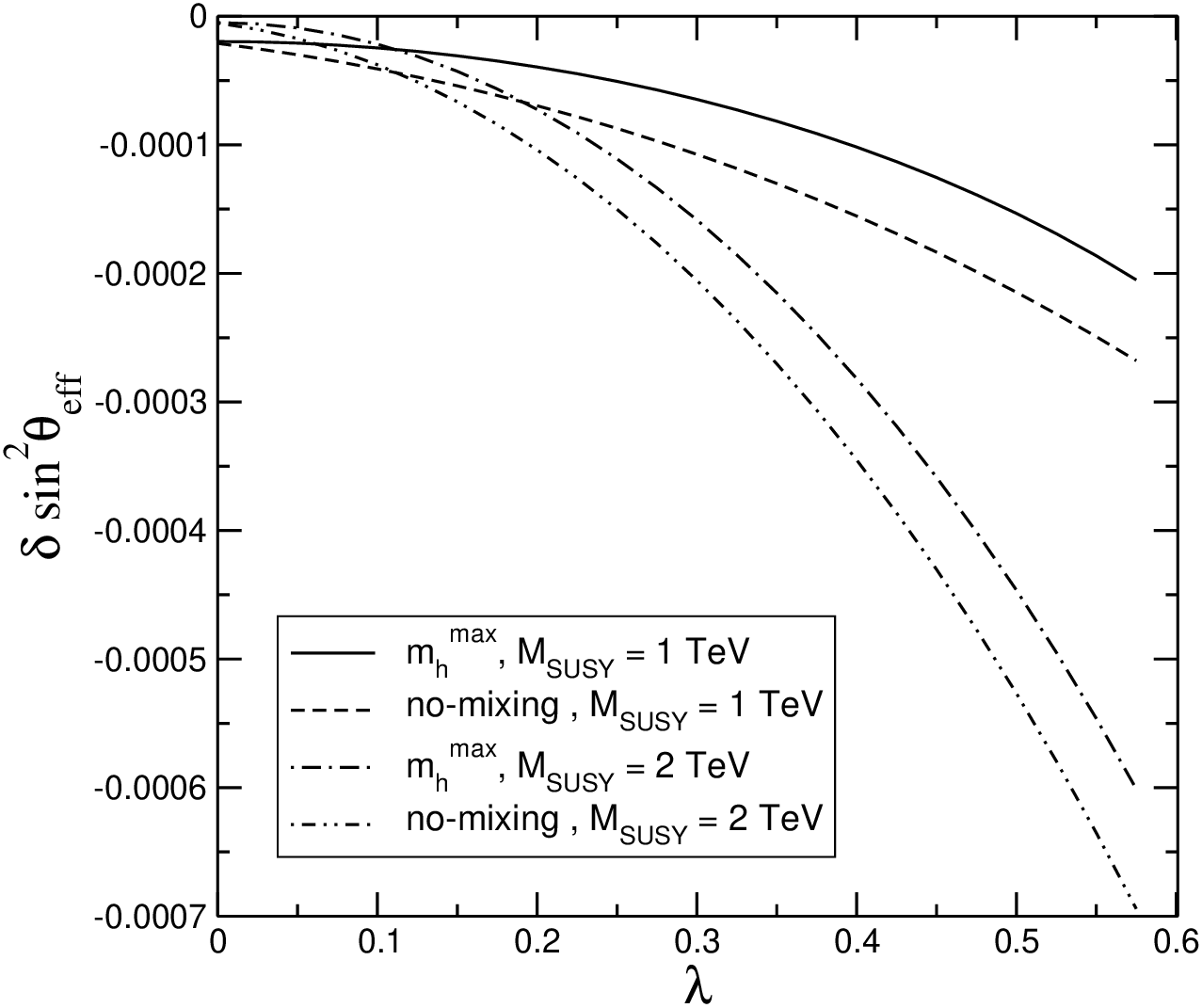,width=7.8cm}
\end{center}\vspace*{-0.6cm}
\caption[]{
The variation of $\de \MW$ and $\de \sweff$ as a function of 
$\la = \la^t = \la^b$, for the $\mhmax$ and
no-mixing~scenarios and different choices of $\msusy$ obtained with
(\ref{precobs}). Using the complete expressions
(\ref{delmw})-(\ref{deltar}) yields practically indistinguishable results.} 
\label{fig:Ewprecision}
\end{figure}

Here the numerical effects of the NMFV contributions on the electroweak
precision observables, $\de \MW$ and $\de \sweff$, are briefly analyzed.
The shifts in $\MW$ and $\sweff$ have been evaluated both from the
complete expressions for the scalar quark contributions, 
eqs.~(\ref{delmw})-(\ref{deltar}), and using the  
$\De\rho^{\tiq}$ approximation (\ref{precobs}). The corrections to 
these two observables based on (\ref{precobs}) as a function of   
$\la (= \la^t = \la^b)$ are presented in \reffi{fig:Ewprecision} with
the other parameters chosen according to~(\ref{eq:inputs}).
The $\mhmax$~scenario and no-mixing~scenario are selected for both
plots, with two values of $\msusy$, as before. 
The induced shifts in $\MW$ can become as large as $0.14 \gev$ for the 
extreme case, i.e.\ when $\msusy=2 \tev$, $\la=0.6$ and the case 
of no-mixing is considered. In the $\mhmax$~scenario $\de\MW$ is
smaller, $\de \MW \lsim 0.05 \gev$, but still sizeable.
Using the complete expressions (\ref{delmw})-(\ref{deltar}) yields
results practically indistinguishable from those shown in
\reffi{fig:Ewprecision}. Thus (\ref{precobs}) is a sufficiently
accurate, simple approximation for  squark-mixing effects in
the electroweak precision observables.

The shifts $\de \sweff$, shown in the right plot of
\reffi{fig:Ewprecision}, can reach values up 
$7 \times 10^{-4}$ for $\msusy=2$ TeV and $\la=0.6$ in the
no-mixing~scenario, being smaller (but still sizeable) 
for the other scenarios chosen here. 

These variations have to be compared with the current experimental
uncertainties~\cite{ewdataw03},
\BE
\Delta \MW^{\rm exp,today} = 34 \mev, \qquad
 \Delta \sweff^{\rm exp,today} = 17 \times 10^{-5} \,,
\end{equation}
and the expected experimental precision for the LHC, 
$\Delta \MW = 15-20 \mev$~\cite{MWatLHC}, and at a future linear
collider running on the $Z$ 
peak and the $WW$ threshold (GigaZ)~\cite{moenig,gigaz,blueband},
\BE
\Delta\MW^{\rm exp,future} = 7 \mev,  \qquad
 \Delta \sweff^{\rm exp,future} = 1.3 \times 10^{-5} ~.
\end{equation}
Extreme parts of the NMFV parameters (especially for $\la^t \neq \la^b$) 
can be excluded already with today's precision. But even small values
of $\la = \la^t = \la^b$ could be probed with the future precision on
$\sweff$, provided that theoretical uncertainties will be sufficiently
under control~\cite{susyewpo}. 

\section{Conclusions}
\label{sec:conclu}

We have calculated the MSSM scalar-quark contributions to electroweak
observables arising from a
NMFV mixing of the third and second generation squarks.
In particular, we have evaluated the lightest MSSM Higgs boson mass, the
$\rho$-parameter, and the 
electroweak precision observables $\MW$ and $\sweff$. 
The analytical results have been obtained for a general 
$4 \times 4$ mixing in the $\Stop/\Scha$ as well as in the
$\Sbot/\Sstr$ sector. They have been included in the Fortran code
\fhto \,(see {\tt www.feynhiggs.de}). 
The numerical analysis has been performed for a
simplified model in which only the left-handed squarks receive an
additional non-CKM mixing contribution.

Numerically we compared the effects of NMFV on the mass of
the lightest MSSM Higgs boson in five benchmark scenarios. For
small and moderate NMFV the effect is small, being at present lower than
the theoretical uncertainty of $\mh$, 
$\de\mh^{\rm theo} \approx 3 \gev$~\cite{mhiggsAEC}. 

We have presented the analytical results for the squark contribution 
to the $\rho$-parameter. The additional contribution can be of
\order{10^{-3}} and can significantly depend on the relative sign of CKM
and non-CKM generation mixing. Even larger contributions can be 
obtained if the mixing in the
$\Stop/\Scha$ and $\Sbot/\Sstr$ sector is varied independently.

Finally we have analyzed the NMFV corrections to the electroweak
precision observables $\MW$ and $\sweff$. We have shown that the 
effects of scalar-quark generation mixing enters essentially through
$\De\rho$. Large parts of the parameter space 
can be excluded already with today's experimental precision of these
observables, and even more for the increasing precision at future
colliders. 

\subsection*{Acknowledgements}
\vspace{-.5em}
We thank T.~Hahn for technical help.
We thank P.~Slavich and M.~Vogt for helpful discussions.
S.H. thanks A.~Dedes, T.~Hurth, S.~Khalil, G.~Moortgat-Pick,
D.~St\"ockinger and G.~Weiglein for interesting discussions.
This work has been supported by the European Community's Human
Potential Programme under contract HPRN-CT-2000-00149 Physics at
Colliders. 
Part of the work of S.P. has been supported by the European Union
under contract No.~MEIF-CT-2003-500030.

\begin{appendix}

\section*{Appendix}

\section{The Feynman rules in the MSSM with NMFV}
In this section we list the Feynman rules for the various vertices
used in this paper. Note that the first generation has been 
completely neglected and the indices have been shifted accordingly:
$m_{u_{1}}$ corresponds to  $m_c$, $m_{u_{2}}$ to  $\mt$, 
$A_1^u$ to $A_c$, $A_2^u$ to $\At$ (and analogous in the for down-type
sector). The CKM matrix, $V_{\rm CKM}$, is defined as in~(\ref{eq:CKM}). 
(The Feynman rules for the general case of three generation mixing can be
obtained by replacing '2' by '3' in the sum and in the $R$~indices.)

\small{
\begin{longtable}{p{0.985\linewidth}}
\feynmancaption{2 Higgs -- 2 Squarks}
$\Mfunction{C}(h,h,\tilde u_{\beta},-\tilde u_{\alpha}) = \Mfunction{-}\frac{\Mfunction{\i}\,e^2\,\sum\limits_{n=1}^{2} 
\left\{\begin{gathered}
\left( \left( c_{2\alpha}\,\MW^2\,s_{\beta}^2 \right) \,\left( -3 + 4\,\sw^2 \right)  + 6\,c_{\alpha}^2\,\cw^2\,m_{u_{n}}^2 \right) \,\left( R^{\alpha,n}_{\tilde u}\,(R^{\beta,n}_{\tilde u})^{*} \right) +\hfill\\
\left( -2\,c_{2\alpha}\,\MW^2\,s_{\beta}^2\,\sw^2 + 3\,c_{\alpha}^2\,\cw^2\,m_{u_{n}}^2 \right) \,\left( 2\,R^{\alpha,2 + n}_{\tilde u}\,(R^{\beta,2 + n}_{\tilde u})^{*} \right) \hfill
\end{gathered}\right\}}{12\,\cw^2\,\MW^2\,s_{\beta}^2\,\sw^2}$\\[5ex]
$\Mfunction{C}(h,h,\tilde d_{\beta},-\tilde d_{\alpha}) = \frac{\Mfunction{\i}\,e^2\,\sum\limits_{n=1}^{2} 
\left\{\begin{gathered}
\left( \left( c_{2\alpha}\,c_{\beta}^2\,\MW^2 \right) \,\left( -3 + 2\,\sw^2 \right)  - 6\,\cw^2\,s_{\alpha}^2\,m_{d_{n}}^2 \right) \,\left( R^{\alpha,n}_{\tilde d}\,(R^{\beta,n}_{\tilde d})^{*} \right) -\hfill\\
\left( c_{2\alpha}\,c_{\beta}^2\,\MW^2\,\sw^2 + 3\,\cw^2\,s_{\alpha}^2\,m_{d_{n}}^2 \right) \,\left( 2\,R^{\alpha,2 + n}_{\tilde d}\,(R^{\beta,2 + n}_{\tilde d})^{*} \right) \hfill
\end{gathered}\right\}}{12\,c_{\beta}^2\,\cw^2\,\MW^2\,\sw^2}$\\[5ex]
$\Mfunction{C}(H,H,\tilde u_{\beta},-\tilde u_{\alpha}) = \Mfunction{-}\frac{\Mfunction{\i}\,e^2\,\sum\limits_{n=1}^{2} 
\left\{\begin{gathered}
\left( \left( c_{2\alpha}\,\MW^2\,s_{\beta}^2 \right) \,\left( 3 - 4\,\sw^2 \right)  + 6\,\cw^2\,s_{\alpha}^2\,m_{u_{n}}^2 \right) \,\left( R^{\alpha,n}_{\tilde u}\,(R^{\beta,n}_{\tilde u})^{*} \right) +\hfill\\
\left( 2\,c_{2\alpha}\,\MW^2\,s_{\beta}^2\,\sw^2 + 3\,\cw^2\,s_{\alpha}^2\,m_{u_{n}}^2 \right) \,\left( 2\,R^{\alpha,2 + n}_{\tilde u}\,(R^{\beta,2 + n}_{\tilde u})^{*} \right) \hfill
\end{gathered}\right\}}{12\,\cw^2\,\MW^2\,s_{\beta}^2\,\sw^2}$\\[5ex]
$\Mfunction{C}(H,H,\tilde d_{\beta},-\tilde d_{\alpha}) = \Mfunction{-}\frac{\Mfunction{\i}\,e^2\,\sum\limits_{n=1}^{2} 
\left\{\begin{gathered}
\left( \left( c_{2\alpha}\,c_{\beta}^2\,\MW^2 \right) \,\left( -3 + 2\,\sw^2 \right)  + 6\,c_{\alpha}^2\,\cw^2\,m_{d_{n}}^2 \right) \,\left( R^{\alpha,d}_{\tilde d}\,(R^{\beta,n}_{\tilde d})^{*} \right) -\hfill\\
\left( c_{2\alpha}\,c_{\beta}^2\,\MW^2\,\sw^2 - 3\,c_{\alpha}^2\,\cw^2\,m_{d_{n}}^2 \right) \,\left( 2\,R^{\alpha,2 + n}_{\tilde d}\,(R^{\beta,2 + n}_{\tilde d})^{*} \right) \hfill
\end{gathered}\right\}}{12\,c_{\beta}^2\,\cw^2\,\MW^2\,\sw^2}$\\[5ex]
$\Mfunction{C}(A,A,\tilde u_{\beta},-\tilde u_{\alpha}) = \Mfunction{-}\frac{\Mfunction{\i}\,e^2\,\sum\limits_{n=1}^{2} 
\left\{\begin{gathered}
\left( \left( c_{2\beta}\,\MW^2\,{t^2_{\be}} \right) \,\left( -3 + 4\,\sw^2 \right)  + 6\,\cw^2\,m_{u_{n}}^2 \right) \,\left( R^{\alpha,n}_{\tilde u}\,(R^{\beta,n}_{\tilde u})^{*} \right) +\hfill\\
\left( -2\,c_{2\beta}\,\MW^2\,\sw^2\,{t^2_{\be}} + 3\,\cw^2\,m_{u_{n}}^2 \right) \,\left( 2\,R^{\alpha,2 + n}_{\tilde u}\,(R^{\beta,2 + n}_{\tilde u})^{*} \right) \hfill
\end{gathered}\right\}}{12\,\cw^2\,\MW^2\,\sw^2\,{t^2_{\be}}}$\\[5ex]
$\Mfunction{C}(A,A,\tilde d_{\beta},-\tilde d_{\alpha}) = \Mfunction{-}\frac{\Mfunction{\i}\,e^2\,\sum\limits_{n=1}^{2} 
\left\{\begin{gathered}
\left( \left( c_{2\beta}\,\MW^2 \right) \,\left( 3 - 2\,\sw^2 \right)  + 6\,\cw^2\,{t^2_{\be}}\,m_{d_{n}}^2 \right) \,\left( R^{\alpha,n}_{\tilde d}\,(R^{\beta,n}_{\tilde d})^{*} \right) +\hfill\\
\left( c_{2\beta}\,\MW^2\,\sw^2 + 3\,\cw^2\,{t^2_{\be}}\,m_{d_{n}}^2 \right) \,\left( 2\,R^{\alpha,2 + n}_{\tilde d}\,(R^{\beta,2 + n}_{\tilde d})^{*} \right) \hfill
\end{gathered}\right\}}{12\,\cw^2\,\MW^2\,\sw^2}$\\[5ex]
$\Mfunction{C}(h,H,\tilde u_{\beta},-\tilde u_{\alpha}) = \Mfunction{-}\frac{\Mfunction{\i}\,e^2\,s_{2\alpha}\,\sum\limits_{n=1}^{2} 
\left\{\begin{gathered}
\left( \left( \MW^2\,s_{\beta}^2 \right) \,\left( -3 + 4\,\sw^2 \right)  + 3\,\cw^2\,m_{u_{n}}^2 \right) \,\left( R^{\alpha,n}_{\tilde u}\,(R^{\beta,n}_{\tilde u})^{*} \right) +\hfill\\
\left( -4\,\MW^2\,s_{\beta}^2\,\sw^2 + 3\,\cw^2\,m_{u_{n}}^2 \right) \,\left( R^{\alpha,2 + n}_{\tilde u}\,(R^{\beta,2 + n}_{\tilde u})^{*} \right) \hfill
\end{gathered}\right\}}{12\,\cw^2\,\MW^2\,s_{\beta}^2\,\sw^2}$\\[5ex]
$\Mfunction{C}(h,H,\tilde d_{\beta},-\tilde d_{\alpha}) = \frac{\Mfunction{\i}\,e^2\,s_{2\alpha}\,\sum\limits_{n=1}^{2} 
\left\{\begin{gathered}
\left( \left( c_{\beta}^2\,\MW^2 \right) \,\left( -3 + 2\,\sw^2 \right)  + 3\,\cw^2\,m_{d_{n}}^2 \right) \,\left( R^{\alpha,n}_{\tilde d}\,(R^{\beta,n}_{\tilde d})^{*} \right) +\hfill\\
\left( -2\,c_{\beta}^2\,\MW^2\,\sw^2 + 3\,\cw^2\,m_{d_{n}}^2 \right) \,\left( R^{\alpha,2 + n}_{\tilde d}\,(R^{\beta,2 + n}_{\tilde d})^{*} \right) \hfill
\end{gathered}\right\}}{12\,c_{\beta}^2\,\cw^2\,\MW^2\,\sw^2}$\\[5ex]
\feynmancaption{2 Squarks -- 2 Gauge Bosons}
$\Mfunction{C}(\tilde u_{\alpha},-\tilde u_{\beta},Z,Z) = \frac{\Mfunction{\i}\,e^2}{18\,\cw^2\,\sw^2}\,\sum\limits_{n=1}^{2}{\left( 3 - 4\,\sw^2 \right) }^2\,R^{\beta,n}_{\tilde u}\,(R^{\alpha,n}_{\tilde u})^{*} + 16\,{\sw}^4\,R^{\beta,2 + n}_{\tilde u}\,(R^{\alpha,2 + n}_{\tilde u})^{*}$\\[5ex]
$\Mfunction{C}(\tilde d_{\alpha},-\tilde d_{\beta},Z,Z) = \frac{\Mfunction{\i}\,e^2}{18\,\cw^2\,\sw^2}\,\sum\limits_{n=1}^{2}{\left( 3 - 2\,\sw^2 \right) }^2\,R^{\beta,n}_{\tilde d}\,(R^{\alpha,n}_{\tilde d})^{*} + 4\,{\sw}^4\,R^{\beta,2 + n}_{\tilde d}\,(R^{\alpha,2 + n}_{\tilde d})^{*}$\\[5ex]
$\Mfunction{C}(\tilde d_{\beta},-\tilde u_{\alpha},Z,-W^{-}) =
\Mfunction{-}\Mfunction{\i}{\,e^2\,\sum\limits_{n,m=1}^{2}V^{n,m}_{\Mvariable{\rm CKM}}\,(R^{\beta,m}_{\tilde d})^{*}\,R^{\alpha,n}_{\tilde u}}{3\,{\sqrt{2}}\,\cw}$\\[5ex]
$\Mfunction{C}(\tilde u_{\alpha},-\tilde u_{\beta},W^{-},-W^{-}) = \frac{\Mfunction{\i}\,e^2}{2\,\sw^2}\,\sum\limits_{n=1}^{2}R^{\beta,n}_{\tilde u}\,(R^{\alpha,n}_{\tilde u})^{*}$\\[5ex]
$\Mfunction{C}(\tilde d_{\alpha},-\tilde d_{\beta},W^{-},-W^{-}) = \frac{\Mfunction{\i}\,e^2}{2\,\sw^2}\,\sum\limits_{n=1}^{2}R^{\beta,n}_{\tilde d}\,(R^{\alpha,n}_{\tilde d})^{*}$\\[5ex]
\feynmancaption{2 Squarks -- Gauge Boson}
$\Mfunction{C}(\tilde u_{\alpha},-\tilde u_{\beta},Z) = \frac{\Mfunction{\i}\,e}{6\,\cw\,\sw}\,\sum\limits_{n=1}^{2}\left( -3 + 4\,\sw^2 \right) \,\left( R^{\beta,n}_{\tilde u}\,(R^{\alpha,n}_{\tilde u})^{*} \right)  + 4\,\sw^2\,R^{\beta,2 + n}_{\tilde u}\,(R^{\alpha,2 + n}_{\tilde u})^{*}$\\[5ex]
$\Mfunction{C}(\tilde d_{\alpha},-\tilde d_{\beta},Z) = \Mfunction{-}\frac{\Mfunction{\i}\,e}{6\,\cw\,\sw}\,\sum\limits_{n=1}^{2}\left( -3 + 2\,\sw^2 \right) \,\left( R^{\beta,n}_{\tilde d}\,(R^{\alpha,n}_{\tilde d})^{*} \right)  + 2\,\sw^2\,R^{\beta,2 + n}_{\tilde d}\,(R^{\alpha,2 + n}_{\tilde d})^{*}$\\[5ex]
$\Mfunction{C}(\tilde u_{\alpha},-\tilde d_{\beta},W^{-}) = \Mfunction{-}\frac{\Mfunction{\i}\,e}{{\sqrt{2}}\,\sw}\,\sum\limits_{n,m=1}^{2}(V^{n,m}_{\Mvariable{\rm CKM}})^{*}\,R^{\beta,m}_{\tilde d}\,(R^{\alpha,n}_{\tilde u})^{*}$\\[5ex]
$\Mfunction{C}(\tilde d_{\beta},-\tilde u_{\alpha},-W^{-}) = \Mfunction{-}\frac{\Mfunction{\i}\,e}{{\sqrt{2}}\,\sw}\,\sum\limits_{n,m=1}^{2}V^{n,m}_{\Mvariable{\rm CKM}}\,(R^{\beta,m}_{\tilde d})^{*}\,R^{\alpha,n}_{\tilde u}$\\[5ex]
\feynmancaption{Higgs -- 2 Squarks}
$\Mfunction{C}(h,\tilde u_{\alpha},-\tilde u_{\beta}) = \Mfunction{-}\frac{\Mfunction{\i}\,e\,\sum\limits_{n=1}^{2} 
\left\{\begin{gathered}
\left( \left( \left( \MW\,\MZ\,s_{\alpha+\beta}\,s_{\beta} \right) \,\left( -3 + 4\,\sw^2 \right)  + 6\,c_{\alpha}\,\cw\,m_{u_{n}}^2 \right) \,R^{\beta,n}_{\tilde u} + \left( c_{\alpha}\,A^{u}_{n} +\right.\right. \\ \left. \left. s_{\alpha}\,\mu^{*} \right) \left( 3\,\cw\,m_{u_{n}}\,R^{\beta,2 + n}_{\tilde u} \right)  \right) \,(R^{\alpha,n}_{\tilde u})^{*}+ \left( \left( \mu\,s_{\alpha} + c_{\alpha}\,A^{u*}_{n} \right) \,\left( 3\,\cw\,m_{u_{n}}\,R^{\beta,n}_{\tilde u} \right) \right.\\ \left.- 4\,\MW\,\MZ\,s_{\alpha+\beta}\,s_{\beta}\,\sw^2\,R^{\beta,2 + n}_{\tilde u} + 6\,c_{\alpha}\,\cw\,m_{u_{n}}^2\,R^{\beta,2 + n}_{\tilde u} \right) \,(R^{\alpha,2 + n}_{\tilde u})^{*}\hfill
\end{gathered}\right\}}{6\,\cw\,\MW\,s_{\beta}\,\sw}$\\[5ex]
$\Mfunction{C}(h,\tilde d_{\alpha},-\tilde d_{\beta}) = \frac{\Mfunction{\i}\,e\,\sum\limits_{n=1}^{2} 
\left\{\begin{gathered}
\left( \left( \left( c_{\beta}\,\MW\,\MZ\,s_{\alpha+\beta} \right) \,\left( -3 + 2\,\sw^2 \right)  + 6\,\cw\,s_{\alpha}\,m_{d_{n}}^2 \right) \,R^{\beta,n}_{\tilde d} + \left( s_{\alpha}\,A^{d}_{n} + c_{\alpha}\,\mu^{*} \right) \right.\\ \left. \left( 3\,\cw\,m_{d_{n}}\,R^{\beta,2 + n}_{\tilde d} \right)  \right) \,(R^{\alpha,n}_{\tilde d})^{*}+
\left( \left( c_{\alpha}\,\mu + s_{\alpha}\,A^{d*}_{n} \right) \,\left( 3\,\cw\,m_{d_{n}}\,R^{\beta,n}_{\tilde d} \right)  -\right. \\ \left. 2\,c_{\beta}\,\MW\,\MZ\,s_{\alpha+\beta}\,\sw^2\,R^{\beta,2 + n}_{\tilde d} + 6\,\cw\,s_{\alpha}\,m_{d_{n}}^2\,R^{\beta,2 + n}_{\tilde d} \right) \,(R^{\alpha,2 + n}_{\tilde d})^{*}\hfill
\end{gathered}\right\}}{6\,c_{\beta}\,\cw\,\MW\,\sw}$\\[5ex]
$\Mfunction{C}(A,\tilde u_{\alpha},-\tilde u_{\beta}) = \Mfunction{-}\frac{\Mfunction{e}\,\sum\limits_{n=1}^{2}\left( -\left( A^{u}_{n} + t_{\beta}\,\mu^{*} \right) \,\left( R^{\beta,2 + n}_{\tilde u}\,(R^{\alpha,n}_{\tilde u})^{*} \right)  + \left( A^{u*}_{n} + \mu\,t_{\beta} \right) \,\left( R^{\beta,n}_{\tilde u}\,(R^{\alpha,2 + n}_{\tilde u})^{*} \right)  \right) \,m_{u_{n}}}{2\,\MW\,\sw\,t_{\beta}}$\\[5ex]
$\Mfunction{C}(A,\tilde d_{\alpha},-\tilde d_{\beta}) = \Mfunction{-}\frac{\Mfunction{e}\,\sum\limits_{n=1}^{2}\left( -\left( \mu^{*} + t_{\beta}\,A^{d}_{n} \right) \,\left( R^{\beta,2 + n}_{\tilde d}\,(R^{\alpha,n}_{\tilde d})^{*} \right)  + \left( \mu + t_{\beta}\,A^{d*}_{n} \right) \,\left( R^{\beta,n}_{\tilde d}\,(R^{\alpha,2 + n}_{\tilde d})^{*} \right)  \right) \,m_{d_{n}}}{2\,\MW\,\sw}$\\[5ex]
$\Mfunction{C}(H,\tilde u_{\alpha},-\tilde u_{\beta}) = \Mfunction{-}\frac{\Mfunction{\i}\,e\,\sum\limits_{n=1}^{2} 
\left\{\begin{gathered}
\left( \left( \left( c_{\alpha+\beta}\,\MW\,\MZ\,s_{\beta} \right) \,\left( 3 - 4\,\sw^2 \right)  + 6\,\cw\,s_{\alpha}\,m_{u_{n}}^2 \right) \,R^{\beta,n}_{\tilde u} + \left( s_{\alpha}\,A^{u}_{n} - c_{\alpha}\,\mu^{*} \right)\right. \\ \left. \left( 3\,\cw\,m_{u_{n}}\,R^{\beta,2 + n}_{\tilde u} \right)  \right) \,(R^{\alpha,n}_{\tilde u})^{*}+
\left( \left( -c_{\alpha}\,\mu + s_{\alpha}\,A^{u*}_{n} \right) \,\left( 3\,\cw\,m_{u_{n}}\,R^{\beta,n}_{\tilde u} \right)  + \right.\\ \left. 4\,c_{\alpha+\beta}\,\MW\,\MZ\,s_{\beta}\,\sw^2\,R^{\beta,2 + n}_{\tilde u} + 6\,\cw\,s_{\alpha}\,m_{u_{n}}^2\,R^{\beta,2 + n}_{\tilde u} \right) \,(R^{\alpha,2 + n}_{\tilde u})^{*}\hfill
\end{gathered}\right\}}{6\,\cw\,\MW\,s_{\beta}\,\sw}$\\[5ex]
$\Mfunction{C}(H,\tilde d_{\alpha},-\tilde d_{\beta}) = \Mfunction{-}\frac{\Mfunction{\i}\,e\,\sum\limits_{n=1}^{2} 
\left\{\begin{gathered}
\left( \left( \left( c_{\alpha+\beta}\,c_{\beta}\,\MW\,\MZ \right) \,\left( -3 + 2\,\sw^2 \right)  + 6\,c_{\alpha}\,\cw\,m_{d_{n}}^2 \right) \,R^{\beta,n}_{\tilde d} + \left( c_{\alpha}\,A^{d}_{n} -\right.\right. \\ \left. \left. s_{\alpha}\,\mu^{*} \right) \left( 3\,\cw\,m_{d_{n}}\,R^{\beta,2 + n}_{\tilde d} \right)  \right) \,(R^{\alpha,n}_{\tilde d})^{*}+
\left( \left( -\mu\,s_{\alpha} + c_{\alpha}\,A^{d*}_{n} \right)\,\left( 3\,\cw\,m_{d_{n}}\,R^{\beta,n}_{\tilde d} \right)\right. \\ \left.  - 2\,c_{\alpha+\beta}\,c_{\beta}\,\MW\,\MZ\,\sw^2\,R^{\beta,2 + n}_{\tilde d} + 6\,c_{\alpha}\,\cw\,m_{d_{n}}^2\,R^{\beta,2 + n}_{\tilde d} \right) \,(R^{\alpha,2 + n}_{\tilde d})^{*}\hfill
\end{gathered}\right\}}{6\,c_{\beta}\,\cw\,\MW\,\sw}$\\[5ex]
\end{longtable}
}

\end{appendix}


\end{document}